| Title | **From Basics to Frontiers: A Comprehensive Review of Plasma-Modified and Plasma-Synthesized Polymer Films** |
|---|---|

| Authors | T. Dufour[1] |
|---|---|
| Affiliations | [1]LPP, Sorbonne Université Univ. Paris 6, CNRS, Ecole Polytech., Univ. Paris-Sud, Observatoire de Paris, Université Paris-Saclay,10 PSL Research University, 4 Place Jussieu, 75252 Paris, France. E-mail: thierry.dufour@sorbonne-universite.fr |

| Summary | This comprehensive review begins by tracing the historical development and progress of cold plasma technology as an innovative approach to polymer engineering. The study emphasizes the versatility of cold plasma derived from a variety of sources including low-pressure glow discharges (e.g., radiofrequency capacitively coupled plasmas) and atmospheric pressure plasmas (e.g., dielectric barrier devices, piezoelectric plasmas). It critically examines key operational parameters such as reduced electric field, pressure, discharge type, gas type and flow rate, substrate temperature, gap, and how these variables affect the properties of the synthesized or modified polymers. This review also discusses the application of cold plasma in polymer surface modification, underscoring how changes in surface properties (e.g., wettability, adhesion, biocompatibility) can be achieved by controlling various surface processes (etching, roughening, crosslinking, functionalization, crystallinity). A detailed examination of Plasma-Enhanced Chemical Vapor Deposition (PECVD) reveals its efficacy in producing thin polymeric films from an array of precursors. Yasuda's models, Rapid Step-Growth Polymerization (RSGP) and Competitive Ablation Polymerization (CAP), are explained as fundamental mechanisms underpinning plasma-assisted deposition and polymerization processes. Then, the wide array of applications of cold plasma technology is explored, from the biomedical field, where it is used in creating smart drug delivery systems and biodegradable polymer implants, to its role in enhancing the performance of membrane-based filtration systems crucial for water purification, gas separation, and energy production. It investigates the potential for improving the properties of bioplastics and the exciting prospects for developing self-healing materials using this technology. |
|---|---|
| Keywords | Plasma polymer devices; plasma polymerization; polymer film growth; adhesion; wettability; crosslinking; surface crystallinity; PECVD; biopolymers |

# 1. Tracing back the roots of plasma processes for polymer applications

While the first techniques for synthesizing polymers or improving their properties date back to the end of the 19th century (Galalith's synthesis in 1897) [1], the advent of plasma processes for this purpose come only a few decades later, in parallel with patents filed in the late 1950s. These patents detail the manufacture of thin polymer films for electronic applications, such as capacitors and transmission lines. In 1959, the Radiation Research Corp patented the first polymerizing gas discharge process [2]. As depicted in **Figure 1a,b**, it involves the production of cold plasma between two electrode-like surfaces of metal substrates, which allow for the deposition of low molecular weight polymer films. In 1960, researchers from the Chemstrand Corporation Research Center built upon this pioneering work and developed a plasma technique for treating synthetic fibers such as nylon and polyethylene terephthalate (PET). Conducted in an argon setting, their technique utilizes ion bombardment to create unique, nano-sized patterns on fiber surfaces, as photographed in **Figure 1c** [3]. The capability of plasma to modify surfaces at the nanoscale was thus highlighted. Meanwhile, other researchers, such as B. J. Split, started exploring similar plasma techniques on different materials such as cellulose fibers, demonstrating the versatility of this technology [4].

A significant shift in the use of cold plasma occurred in 1965 when H. F. Sterling and C. G. Swann innovated the Plasma-Enhanced Chemical Vapor Deposition (PECVD) technique. The PECVD process, which synthesizes several micrometer-thick amorphous silicon layers from silane gas in a low-pressure RF discharge, highlights the potential of plasma technology in the field of semiconductor manufacturing [5]. Two years later, Bell Telephone Laboratories adapted their radiofrequency coil device for another purpose: removing hydrogen and fluorine atoms from polyethylene (PE) and polytetrafluoroethylene (PTFE) samples [6], proving plasma's efficiency in chemical treatments.

The field of plasma technology continued to evolve in 1968 when, for the first time, C. Y. Kimand D. A. I. Goring applied corona discharge to roughen the surface of polyethylene, hence demonstrating the possibility of modifying polymer surface topography [7]. The same year, a new patent was filed where monomer vapors were used to increase polymerization rates in a glow discharge [8]. This method not only reinforced the connection between plasma technology and polymer science, but also introduced an avenue to improve process efficiency. Closing out the decade, R. Hollahan (Boeing's Scientific Research Laboratories) demonstrated the ability of plasma technology to chemically modify polymer surfaces, through the utilization of $NH_4$ plasma to functionalize polyethylene with amine groups [9,10]. This landmark achievement marked a milestone in the decade's progress, verifying that plasma technology was a powerful tool for specific chemical modifications of surfaces.

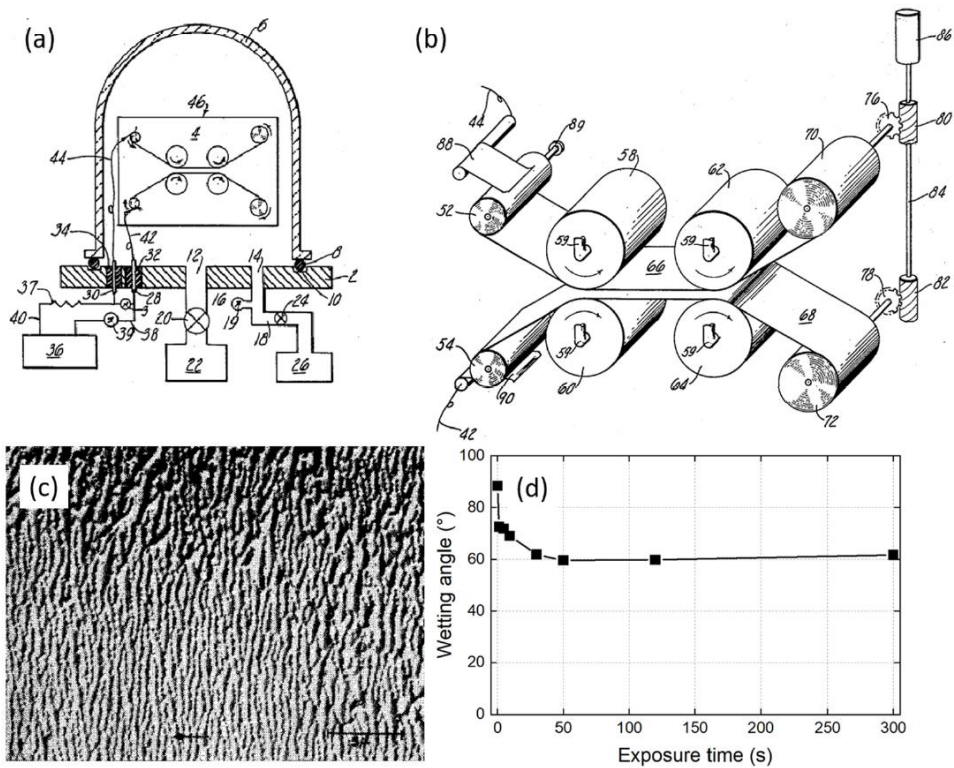

*Figure 1. (a) View of a partial cross-section of an apparatus suitable for carrying out the teachings of the invention; (b) Enlarged view of the substrate transport system utilized in the apparatus of (a). In a vacuum chamber "6", a high voltage is applied by the generator "36" to an electrical wire "44" connected to a flexible contact strip "88" in contact with the supply spool "52" while the ground is imposed via "42" and "90" to the other supply spool "54". Two flexible and metal substrates "66", "68" are unwound from supply spools "52", "54" and pass along convergent paths to guide rollers "58", "60" [2]. (c) SEM micrograph showing the etched surface of a Nylon fiber after 5 min of plasma treatment using the process of Anderson in 1960 (reproduced with permission) [3] Copyright 2004 AIP Publishing. (d) Effect of corona duration on nylon peel strength and polypropylene wettability, adapted from [11].*

The 1970s were characterized by the sporadic publication of original research articles in which surface characterization tools were tested for the first time on plasma-modified/synthesized polymers. As early as 1971, Blais et al. conducted water contact angle (WCA) measurements on polypropylene (PP) samples after their exposure to a $N_2$ corona discharge [11]. The authors noted a decrease in WCA from 85° to 60°, as evidenced in **Figure 1d**. The same article introduced the use of attenuated total reflection (ATR) infrared (IR) spectroscopy to evidence C-C crosslinking and unsaturation. A few years later, X-ray photoelectron spectroscopy (XPS) was utilized to investigate polymer films deposited by the glow discharge polymerization of tetrafluoroethylene, 1,1 difluoroethylene and chlorotrifluoroethylene [12], but also to analyze the elemental surface composition of plasma-treated polymers such as polystyrene (PS), cellulose and nylon [13]. Complementarily to XPS, Auger electron spectroscopy (AES) was proposed by Kny et al., owing to two additional advantages: a higher spatial resolution (detailed surface composition analysis) and a higher surface sensitivity (due to its smaller probing depth) [14]. These analytical techniques are precious to decipher the first initial mechanisms of surface modification and growth of polymer films from a fundamental research perspective [15, 16, 17]. The late 1970s marked the emergence of secondary ion mass spectrometry (SIMS), immediately recognized as a potentially powerful tool for characterizing polymers subjected to plasma sources such as corona discharges [18].

Throughout the 1970s and 1980s, research was primarily centered on the fundamental and practical issues surrounding polymer surface energy and, therefore, on the concepts of wettability and adhesion. The initial techniques in low-pressure plasma are devised through the comparison of diverse gases such as argon, oxygen, nitrogen, air, carbon dioxide and ammonia, aiming to enhance the hydrophilicity of polymer surfaces such as high-density polyethylene (HDPE), RTV silicone and PET [19, 20]. The ability of plasma processes to increase the hydrophobic character of polymers is also demonstrated through ATR-IR and XPS analyses on fluorocarbon or silyl-amine plasma coatings [21, 22]. While wettability describes the extent to which a liquid can spread over a solid surface, adhesion is considered a different concept: it refers to the ability of two different materials to bond, such as parylene polymer on glass or on PP substrate [23]. Good wettability often promotes adhesion without necessarily guaranteeing it, because successful adhesion also depends on strong intermolecular interactions between the materials [24].

In parallel to the booming of research on surface wettability and adhesion, the late 1970s was also marked by the development of plasma processes that paved the way for microelectronics manufacturing. In 1979, Yasuda proposed the competitive ablation and polymerization (CAP) mechanism likely to occur in glow plasma processes supplied in CF4 with/without $H_2$ [25]. While polymer deposition and etching mechanisms were also investigated in various gas mixtures (e.g., $C_2F_6$, $O_2$-$CF_4$, ...) in low-pressure RF discharges [26], it is only a few years later that the coexistence of these two mechanisms is unambiguously demonstrated by Kitamura et al. when exposing $SiO_2$ substrates to a $C_2F_6$ radiofrequency (RF) plasma [27]. As shown in **Figure 2a**, the authors evidence that the film thickness of the deposited fluorocarbon polymer increases with the plasma exposure time while the underlying $SiO_2$ layer is simultaneously etched over an increasing depth (non-linear profile). These results cause the authors to conclude that the growth of the deposited film inhibits the etching reaction.

As an alternative to RF excitation, microwave plasmas supplied in $CF_4$, $SF_6$, $O_2$-$CF_4$ or $O_2$-$N_2$ have been investigated owing to their higher plasma densities responsible for a more effective polymer etching (e.g., photoresist and polyimide samples) [28, 29, 30]. From the late 1980s to now, plasma etching has considerably improved to meet important processing requirements, such as etching rate, selectivity, directionality, and minimization of contamination by metals or polymers, among other factors; all of which have greatly facilitated the advancement of device miniaturization [31].

In 1992, the first atomic force microscopy (AFM) characterizations were carried out on plasma-treated/deposited polymers (e.g., polyimide, hydrocarbon polymers) [32, 33]. This technique clearly evidences how plasma can roughen and texturize a polymer surface into different geometrical patterns [34]. Simultaneously, with the expanding advancements in wettability/adhesion and microelectronics, the 1990s were distinguished by a growing body of literature focused on cleaning and biocompatibility. It was found that plasma cleaning could effectively remove organic contaminants from polymer surfaces such as Si (100) wafers [35]. While some processes combine plasma etching of the uppermost $SiO_2$ layer followed by plasma cleaning to remove the C-F polymeric residues [35], other processes combine a wet cleaning approach with hydrogen plasma to passivate these wafers [36]. Hence, any residual carbon and/or oxygen are removed by attaching a layer of H atoms. A comprehensive review was published as early as 1993 on this subject, deciphering cleaning, ablation, crosslinking and surface chemical modification [37].

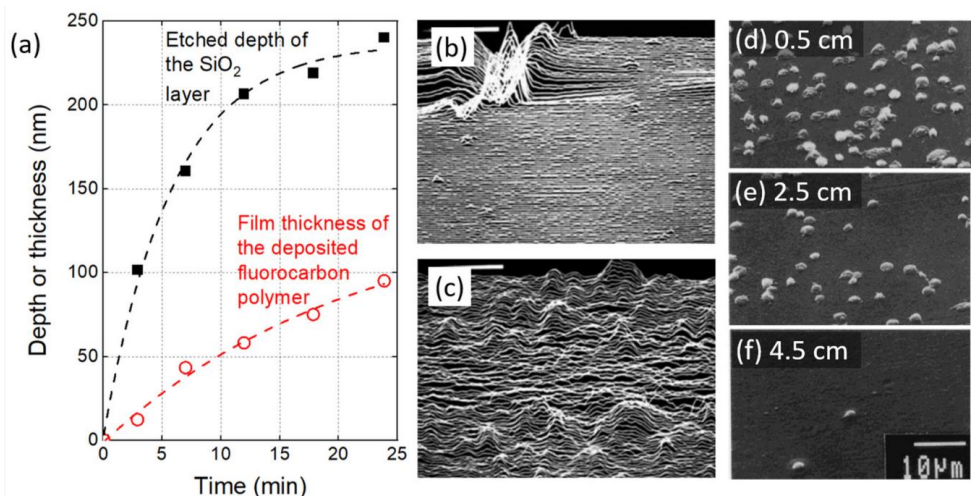

*Figure 2. (a) Time variations in the thickness of a fluorocarbon polymer film deposited on a $SiO_2$ layer and time variation in the etched thickness of the $SiO_2$ layer, adapted from [27]. (b) AFM picture of native polyimide surface and (c) of the same sample after plasma exposure (reproduced with permission) [32] Copyright (c) 1992 John Wiley & Sons, Inc. (d–f) SEM pictures of platelets adhered on PEO10-MA-grafted PE surface at different positions (0.5, 2.5, 4.5 cm) from the untreated end of the gradient surface (reproduced with permission) [38] Copyright (c) 1992 John Wiley & Sons, Inc.*

By modifying the surface characteristics of polymers, plasma treatments can also improve cell adhesion, proliferation and overall biocompatibility, which corresponds to the second emerging research area of this decade. For instance, functional groups can be introduced onto polyethylene oxide surfaces to enhance protein adsorption, which in turn promotes platelet adhesion, as evidenced in **Figure 2d-f** for different treatment positions [38]. Furthermore, plasma treatments can also modify the polymer surface roughness or topography, thus influencing cell behavior, as observed in the case of fluorocarbon polymers (PTFE, FEP) [39] and plasma-deposited polysulphone and poly(hydroxybutyrate) membranes for a bioartificial pancreas device [40].

The 2000s were marked by an exponential increase in the previous areas of interest (wettability/adhesion, advancing microelectronics, biocompatibility, cleaning, etc.). Specialized sectors emerged from these thematic to meet specific challenges. Notable examples include the development of dedicated processes for sterilizing medical devices [41, 42, 43, 44], the creation of innovative self-cleaning surfaces, anti-fog and anti-icing coatings [45, 46, 47, 48] and meeting the requirements of the food packaging industry through the synthesis of oxygen barriers, moisture barriers and gas barriers [49, 50, 51, 52]. In a context marked by sustainable development policies, biopolymers (i.e., polymers produced by living organisms such as cellulose, chitosan, starch and proteins) represent a sustainable and environmentally friendly alternative with substantial potential for film development. Plasma processes were investigated to improve their wettability properties as well as their thermo-mechanical and barrier properties [53, 54].

Since the 2010s, plasma processes have been increasingly specialized and diversified. Nanoscale film deposition and surface modification techniques have been developed, while emphasis has been placed on environmentally friendly plasma processes. The development of atmospheric-pressure plasma systems has also enabled the treatment of heat-sensitive materials, thus increasing the practicality of plasma processes in industries such as textiles and packaging. Even if the potential of cold plasma has already been widely exploited, ongoing technological advances point to an exciting future for plasma processes in polymer processing and film growth, as detailed in **Section 5**.

# 2. Cold plasma generation for the treatment of polymers

## 2.1. Principle of cold plasma generation and main properties

Cold plasma, also known as non-thermal plasma, represents a partially ionized gas state where less than 1% of the gas molecules are ionized, limiting heat generation [55]. The electron temperature far exceeds the ion temperature, which in turn exceeds the gas temperature, typically by tens to hundreds °C [56]. The absence of thermal equilibrium within this gaseous medium explains the terminology of "cold plasma", as opposed to the conditions encountered in a fully ionized plasma, such as in stars or fusion reactors, to which the term "hot plasma" is therefore attributed.

Cold plasmas are characterized by electrical, chemical, radiative, thermal and fluid-mechanical properties, as sketched in **Figure 3a**. In general, cold plasma generation involves the existence of an electric field obtained by applying a high voltage between two electrodes in a gas-filled chamber, either at atmospheric or reduced pressure. High-energy electrons can be created and then collide with the gas molecules to produce an array of active species, including ions, free radicals, excited molecules and ultraviolet photons [57]. As depicted in **Figure 3b**, vacuum ultraviolet (VUV) radiation demonstrates a significant penetration depth in PET polymer, typically reaching up to 100 nm. In contrast, both positive ions and neutral radicals primarily reside within the uppermost surface layers, exhibiting a typical penetration depth of approximately 1 nm [58]. These different species significantly contribute to the modification of polymer surfaces or the synthesis of polymer films onto substrates (see **Section 2.3**).

## 2.2. Cold plasma devices for polymer treatment and film growth

Various methods are available to generate cold plasma, including DC, RF and microwave discharges, each with its own advantages and disadvantages depending on the application.

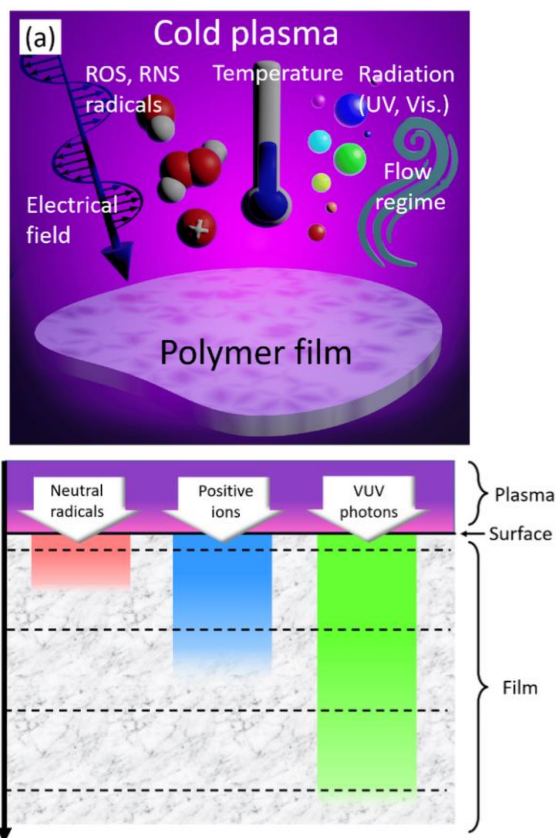

**Figure 3.** (a) Diagram representing a polymer film subjected to cold plasma, highlighting its key properties. (b) Illustration of the penetration depth for neutral radicals, positively charged ions and VUV radiation within PET. Adapted from [58].

### 2.2.1. DC glow plasma devices

Direct Current (DC) plasmas predominantly operate at low pressure, within a few milliTorr to Torr, to prevent the glow-to-arc transition. The discharge is initiated by applying a high voltage between two electrodes, typically in the range of hundreds to thousands of volts. Under the influence of the resulting electric field, positive ions in the plasma are accelerated towards the negatively biased electrode (the cathode) with the ability to sputter its surface. Depending on the process conditions and the choice of process gas, DC glow plasma can etch polymer surfaces (e.g., polysulfone membranes) [59] or deposit polymer coatings onto substrates (e.g., hexamethyldisiloxane plasma polymerization) [60]. Two types of DC glow plasmas can be utilized to process polymers: DC continuous discharges and DC pulsed discharges.

As an example of DC continuous plasma, **Figure 4a** shows a coaxial reactor composed of a glass tube surrounded by two ring electrodes, one powered to a DC voltage of 700 V, the other grounded [60]. The vector gas and the monomer vapor are injected in the interelectrode region (<1 cm³), which provides a stable flow of the plasma-monomer mixture. As ion bombardment causes the cathode to heat up during the process, cooling systems may be required for extended operating times. An alternative for reducing the heating is to replace continuous DC with pulsed DC voltage, where frequency (typically in the 100 Hz–13.56 MHz range) and duty cycle play critical roles in the modified/synthesized polymer films [61, 62]. **Figure 4b** represents a DC pulsed plasma device where a polymer film is modified/grown on the grounded electrode in a confined low-pressure reactor [59, 61]. Each periodic pulse can be divided into two distinct phases: the "pulse on" phase, where the fragmentation and recombination of monomers drive polymerization, and the "pulse off" phase, where the free radicals, generated during the pulse on phase, interact with monomers to set off the polymerization process [63].

### 2.2.2. Dielectric barrier devices

Dielectric barrier devices (DBDs) offer simplified design and operation for the treatment of polymers. Regardless of their specific configurations [64], all DBD fundamentally consist of two electrodes (a powered one and a grounded one) separated by at least one dielectric barrier, as depicted in **Figure 4c**. Operation requires high voltages, typically in the kilovolt range, coupled with relatively low currents [65]. Alternative current (AC) is critical to this process because it allows for the repeated formation and extinction of plasma in each half-cycle. As the voltage changes direction, the electric field reverses, leading to the discharge being quenched (extinguished) and then re-initiated [66]. This pattern creates a sequence of very brief micro-discharges that generate the plasma, while the intervening quenching prevents excessive heating of the gas and transition to an arc discharge. If DBD can treat polymers at atmospheric pressure (**Figure 4c**), it can also be placed within a vacuum chamber (**Figure 4d**) to work in specific gas environments (such as helium, argon and nitrogen) either at lower pressures or in a post-purged atmospheric pressure. The selection of the treatment gas allows for the generation of various reactive species and obtains a wide spectrum of surface modifications. Although the intermittent nature of microdischarges can lead to a non-uniform plasma, careful DBD design and operation can effectively mitigate this issue [67].

Plasma deposition of thin polymer films can be achieved by coupling the DBD with a bubbler to store a liquid monomer, as sketched in **Figure 4d** [68]. An inert carrier gas (often helium, argon or nitrogen) is introduced at the bottom of the bubbler. As the gas travels upwards, it comes into contact with the liquid monomer and bubbles through it, causing its evaporation and mixture with the carrier gas. Then, this mixture of carrier gas and monomer vapor is delivered into the DBD reactor where plasma polymerization takes place. The high-energy electrons in the plasma can break the monomer molecules apart, allowing them to recombine and finally form a polymer layer [68, 69]. The amount of monomer in the carrier gas (and hence, delivered to the reactor) can be controlled by adjusting the flow rate of the carrier gas and the temperature of the bubbler. A higher flow rate or temperature will result in more monomers being delivered and, therefore, a higher deposition rate, as demonstrated using acrylic acid monomer [70] or liquid propylene [71].

### 2.2.3. Radio frequency (RF) plasma devices

Radio frequency capacitively coupled plasmas (RF-CCP) generally operate at low pressure (from a few mTorr to ten Torr) with an ISM (Industrial, Scientific and Medical) frequency of 13.56 MHz. As represented in **Figure 4e**, the RF voltage is applied to an electrode, hence generating an oscillating electric field that drives the plasma toward the counter-electrode [72]. The plasma behaves as a dielectric the impedance of which can change with operating conditions such as power, pressure and gas composition. When the plasma impedance does not match the impedance of the RF generator, some of the power is reflected to the generator. As a result, the RF generator risks being damaged while inefficient power is transferred to the plasma. To prevent this situation, a matching network is designed to adjust the impedance (seen by the RF generator) to the plasma impedance [73]. This device is typically composed of inductors utilized to build up and store energy, as well as variable capacitors to tune the impedance and create a match between the source and the load. Some matching networks may use a "Π", "L" or "T" configuration, depending on the needs of the application.

When the RF voltage is applied to the plasma, the electrons move much more quickly toward the powered electrode during the positive half-cycle than the heavier ions do during the negative half-cycle. This results in a net positive charge on the electrode. To balance the charge in the system, a negative direct current (DC) voltage, or self-bias, builds up on the RF electrode (**Figure 4f**) [74]. This self-bias attracts positive ions from the plasma toward the electrode. In response to this process, a region known as a 'plasma sheath' forms near the electrode. This sheath has a lower plasma density than the bulk plasma and serves to accelerate the ions towards the electrode due to the electric field within it. The sheath oscillates with the RF cycle, expanding and contracting, which modulates the energy of the ions impacting the electrode and consequently affecting etching or deposition rates. Given the selection of suitable process gases and conditions, RF-CCP exhibits versatility in applications, ranging from surface modification to thin film growth [75, 76].

As a technological variation, a dual RF CCP process is designed to independently control ion energy and plasma density. **Figure 4g** represents this process, where: (i) one electrode operates at high frequency (13.56 MHz or 27.12 MHz), resulting in a low-voltage and low-energy sheath, which facilitates the control of ion density and, therefore, the control of deposition and surface modification reactions [78]; (ii) the other electrode is polarized at lower frequency (e.g., 2 MHz), hence generating a high sheath voltage, which subsequently provides high ion energy that is especially beneficial for physical sputtering or etching [79]. The ability to independently manipulate etching and deposition rates makes dual RF CCP an efficient and versatile process for semiconductor manufacturing and surface modification applications.

Complementarily to RF (dual)-CCP, radio frequency inductively coupled plasma (RF-ICP) is an effective method for generating high-density plasmas, typically $10^{10}$-$10^{11}$ cm$^{-3}$ [80]. It operates at similar pressure ranges and frequencies as RF-CCP, but features an inductive coil, usually a flat spiral located above the chamber or surrounding it, as shown in **Figure 4h**. This coil, which carries the RF current, creates an oscillating magnetic field that induces an electric field in the gas, causing its ionization. The main advantage of RF-ICP is its ability to generate plasmas with densities often an order of magnitude higher than those of RF-CCP sources [80], making it an ideal choice for high-speed etching or deposition processes in surface modification and thin film growth [81].

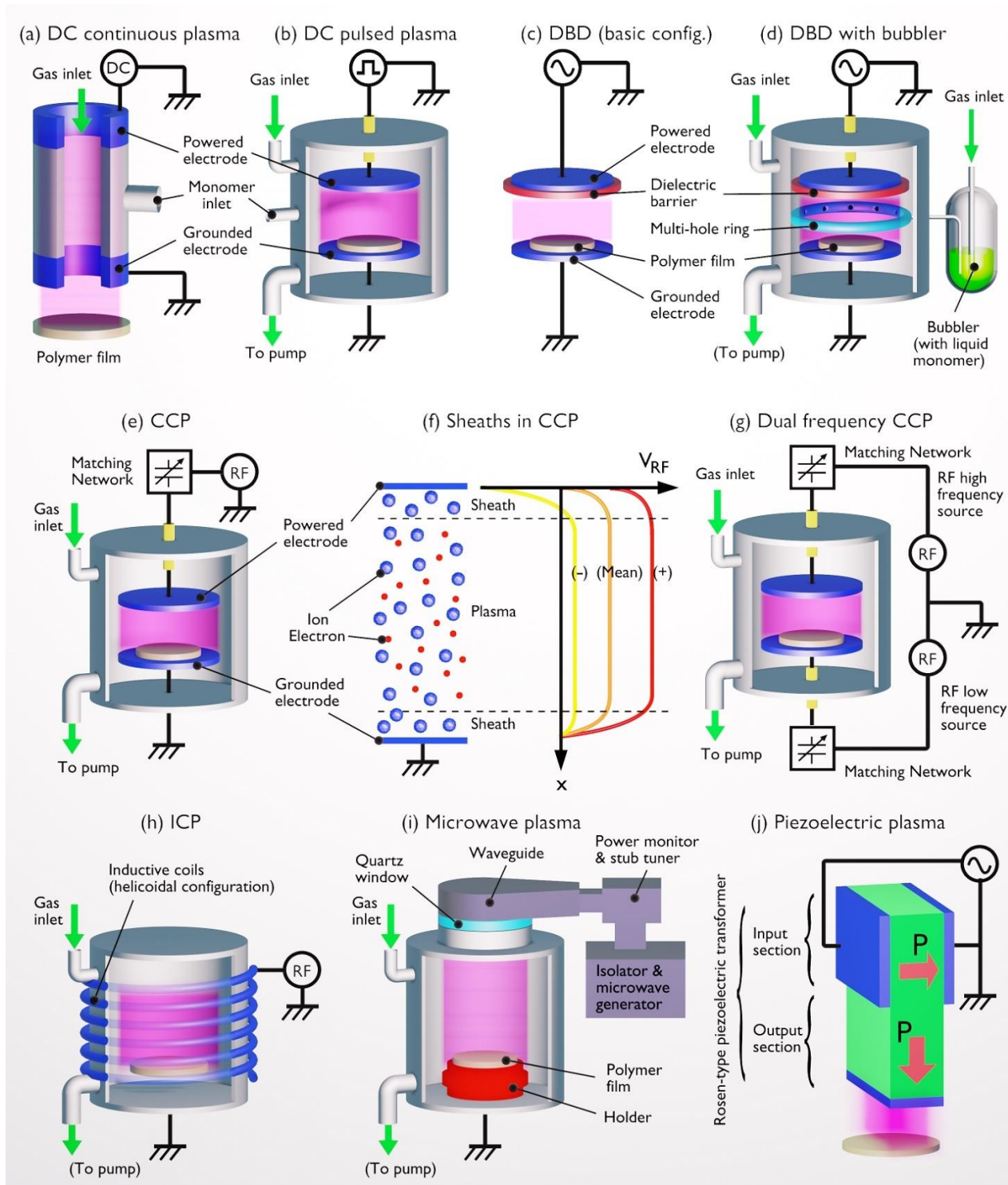

**Figure 4.** Depiction of routinely employed plasma sources for the modification or the synthesis of polymer films. (a) Low-pressure reactor supplied with DC voltage; (b) Low-pressure reactor supplied with DC pulsed voltage; (c) Dielectric barrier device (DBD) operating in ambient air; (d) DBD operating in a low-pressure reactor supplied in monomer vapor by a bubbler; (e) Capacitively coupled plasma (CCP) device supplied with RF voltage; (f) Diagram correlating the position of the ionic sheaths and the plasma region with the axial profile of the RF voltage for maximum (+), minimum (−) and average values; (g) Dual frequency CCP; (h) Inductively coupled plasma (ICP) device; (i) Microwave plasma device; (j) Piezoelectric plasma device with Rosen-type transformer.

### 2.2.4. Microwave plasma devices

In microwave plasma devices, electromagnetic waves are typically generated at 2.45 GHz or 915 MHz, which are standard frequencies for industrial microwave sources [82]. These waves are then directed through an isolator, as illustrated in **Figure 4i**. The isolator ensures that the microwaves only flow in a single direction, safeguarding the microwave generator from potential damage caused by power reflection. As these microwaves propagate within the system, their power is controlled by a monitor and can be adjusted before they encounter a stub tuner. This device matches the impedance between the microwave source and the plasma load for maximizing power transfer and minimizing reflected power. Finally, the microwaves are directed by a waveguide (or a microwave cavity) into the plasma chamber where they interact with the gas molecules to ionize them and create a plasma.

Microwave plasmas are generally generated in an electrodeless configuration, a feature that protects polymers from potential contamination from electrode materials [83]. Depending on system design and application, microwave plasmas can operate in a wide range of pressures, from vacuum to atmospheric pressure. In some configurations, a solenoid can be coaxially centered around the plasma chamber, similar to the one sketched in **Figure 4i**. This solenoid creates a magnetic field that drives electrons into helical trajectories, increasing their path length and thus the probability of ionizing collisions [84]. Electron cyclotron resonance (ECR) can be attained if the microwave frequency matches the natural frequency of electron gyration in the magnetic field. In this case, the electrons absorb energy from the microwave field and acquire significant energy. This high-energy electron population can then ionize the gas more effectively, creating a high-density plasma. The ECR-based microwave plasma devices are advantageous because of their ability to produce high-density plasmas at low pressure. They can operate at lower pressures than other types of plasma sources, which can be beneficial for processes such as highly selective etching [85] or film deposition, where low-pressure operation allows greater control and uniformity.

### 2.2.5. Piezoelectric direct discharge devices

Piezoelectric direct discharge (PDD) devices represent a recent development in the creation of cold plasma under atmospheric pressure conditions using gaseous discharge. A PDD is based on a Rosen-type piezoelectric transformer (PT), which consists of two sections represented in **Figure 4j**: an input section (the primary) and an output section (the secondary), both made of a single piece of piezoelectric material, often lead zirconate titanate [86]. The primary and secondary are separated by a nodal line where vibration amplitude is minimal.

During operation, an alternative voltage is applied to the primary section, typically with a magnitude of 12 or 24 V. To ensure that the PT operates with maximum efficiency, the frequency of this input voltage is selected to align with the PT resonance frequency, in the range of tens to hundreds of kilohertz [87]. This input voltage causes the PT to vibrate due to the inverse piezoelectric effect (electric field causing mechanical deformation). Then, these vibrations cause an alternating electric field in the secondary section due to the direct piezoelectric effect (mechanical deformation causing an electric field). As a result, a high voltage is produced in output, which ionizes the ambient air, leading to the formation of a cold plasma.

PDDs have been successfully applied in controlling the surface free energy of polymers such as HDPE and poly (methyl methacrylate) (PMMA) [88]. Furthermore, they can improve the biocompatibility of orthopedic implants, such as GUR 1020 polymer [89]. Indeed, the biocompatibility characteristics of these plasma-coated samples are enhanced by a factor of 2 to 3, while tribological wear rates are reduced by a factor of 60.

## 2.3. Key plasma parameters affecting polymer treatment and film growth

The choice of plasma characteristics depends on the specific application requirements and the plasma conditions need to be carefully optimized to achieve the desired surface properties or film thickness. This section introduces the main plasma characteristics that affect polymer treatment and film growth.

### 2.3.1. Reduced electric field

The reduced electric field, also known as the electric field strength (E/n), characterizes the electric field (E) in the plasma relative to the number density of neutral particles (n). It is usually measured in Townsend (Td), where 1 Td = $10^{-21}$ V.m$^2$. The E/n parameter controls both volume and surface reactions:
- It determines the rates of various processes occurring in the gas phase (ionization, attachment or excitation) which are crucial in initiating and sustaining the reactions that lead to polymer film growth or modification [90];
- It impacts the energy of ions and radicals reaching the substrate surface, which in turn can affect film properties such as density, composition and bonding structure. For instance, a high E/N might lead to more fragmentation of precursor molecules and the formation of films with different chemical and physical properties.

In an equimolar $N_2$-$O_2$ plasma, electron collisions with nitrogen or oxygen molecules generate reactive species such as N, O, NO, $N_2^+$ and $O_2^+$. Each type of collisional process, whether excitation, ionization or dissociation, results in a loss of a certain fraction of electron energy that directly depends on the applied E/N value [91]. Thus, of the eight collisional processes introduced in **Figure 5**, $N_2$ and $O_2$ ionizations (most prominent at 1000 Td), $O_2$ dissociation (100 Td) and electronic excitation of $O_2$ (300 Td) are typically most efficient in generating reactive oxygen species (ROS). These ROS are likely to catalyze more powerful surface effects such as etching, crosslinking and oxidation. While vibrational and rotational excitations also occur in plasma, these processes deal with lower energy states and are less likely to directly contribute to ROS formation [92]. Often, these vibrational and rotational energy states only serve as intermediate steps in the energy cascade from electronic excitation to thermal equilibrium. For these reasons, a DBD (typically > 100 Td) is more relevant than a glidarc (typically 10-100 Td), for which the dominant process is the vibrational excitation of $N_2$.

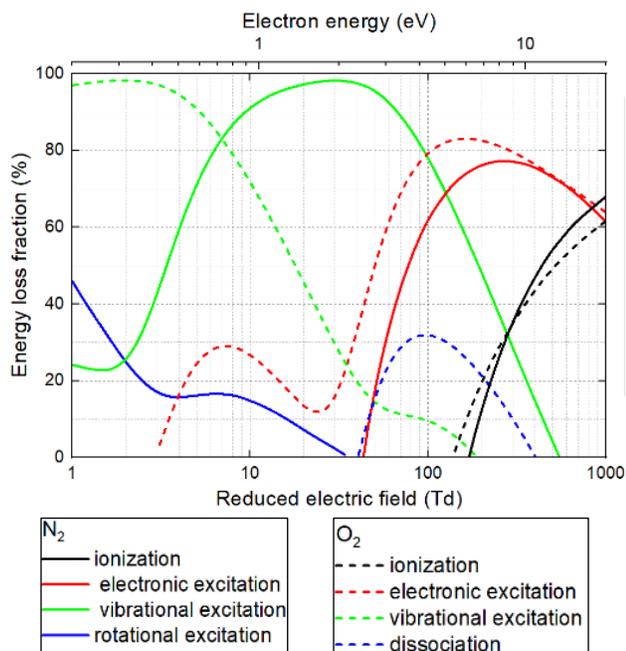

Figure 5. *Proportion of electron energy allocated to different channels of excitation, ionization and dissociation of $N_2$ and $O_2$ within a $N_2$-$O_2$ (50–50%) plasma. This energy loss fraction is represented as a function of the reduced electric field (E/N), computed based on the associated cross-sections of the electron impact reactions. Adapted from [91].*

### 2.3.2. Plasma gas composition

Distinct gases display unique responses to the same electric field, or E/N value, given their specific dissociation, metastable and ionization energies, as outlined in **Table 1**. Noble gases such as helium, neon and argon are frequently utilized as carrier gases due to their inherent chemical stability, enabling the creation of controlled conditions. Significantly, their high-energy metastable states, such as the 19.8 eV state of helium, can transfer energy to other gas species through processes such as the Penning ionization of $N_2$ or $O_2$. These metastable species do not respond to the electric field and can also proceed to polymer modifications.

Table 1. *Dissociation, metastable and ionization energy for noble and diatomic gases [93].*

| Gas | Dissociation Energy (eV) | Metastable Energy (eV) | Ionization Energy (eV) |
|---|---|---|---|
| He | | 19.8 | 24.6 |
| Ne | | 16.6 | 21.6 |
| Ar | | 11.5 | 15.8 |
| $H_2$ | 4.5 | | 15.6 |
| $N_2$ | 9.8 | | 15.5 |
| $O_2$ | 5.1 | | 12.5 |

The energetic levels reported in **Table 1** are responsible for all the reactive chemistry and physical processes involved in surface modification and film growth processes. Depending on the desired outcome, different plasma gases or mixtures are commonly utilized:

- $O_2$ plasmas increase the levels of oxygen-containing functional groups such as hydroxyl (-OH), carbonyl (-C=O) and carboxyl (-COOH) to improve the wettability of polymers such as PE, polyurethane, vinyl-trimethoxysilane-grafted ethylene-propylene and PMMA [94, 95, 96, 97]. Interestingly, oxygen plasma can also selectively erode the organic component of polydimethylsiloxane (PDMSO)-like films applied to hydrophobic polymers, leading to the creation of a hydrophilic surface [98].
- $N_2$ plasmas incorporate nitrogen atoms into polymer surfaces while ensuring lower oxidation [99] and increasing $sp^2/sp^3$ ratios of bonded carbon atoms [100]. Nitrogen plasmas find applications in the microelectronics industry (printed circuit boards, flexible displays and sensors) [101, 102], in the food packaging industry (e.g., enhancing the adhesion of polymer films to metals and glass) [103] and in biomedical applications (improving cell adhesion and proliferation on PLLA scaffolds [104], increasing the self-bonding strength of PEEK surfaces [105]).
- $H_2$ plasmas can increase the roughness of PE surfaces without necessarily enhancing adhesion or surface chemistry, unlike $O_2$ and $N_2$ plasmas [106]. Pulsed DC PECVD supplied in $H_2$ can also selectively etch the outer interface between hydrogenated amorphous carbon (a-C:H) films and silicon layers to improve adhesion below 300 °C [107]. More generally, hydrogen plasma can facilitate the exploration of how hydrogen ions, atoms and UV radiation interplay with polymer surfaces, such as PET [108].
- Thanks to its reactive triple bond, acetylene ($C_2H_2$) is a key precursor gas in PECVD to synthesize amorphous carbon layers (a-C:H) with diamond-like properties (high hardness, chemical inertness, high electrical resistance) [109]. When combined with maleic anhydride in plasma copolymerization, it helps form gentamycin-loaded nanofibers that exhibit antibacterial and biocompatible properties for wound healing [110]. Additionally, acetylene can be decomposed to form carbon nanocoatings on polyurethane surfaces, facilitating their wettability [111]. The resulting films are suitable for flexible and stress-resilient bioimplants.
- Water vapor ($H_2O$) admixed to a carrier gas such as argon generates OH and H radicals, capable of efficiently etching polystyrene (PS) [112] or inducing roughness on PMMA surfaces while integrating hydrophilic O-C=O groups [113]. Admixing water vapor to an Ar-$NH_3$ plasma introduces other oxygen-containing groups, facilitating the deposition of a polydopamine/polyethyleneimide layer [114].
- $CO_2$ plasmas can introduce highly negative carboxylate (COO-) groups on PVDF membranes, hence improving the adsorption of toxic crystal violet dye and iron oxide nanoparticles from water [115].
- $CO_2$-$C_2H_4$ plasmas in an argon environment can be employed as a polymerization process on nanofibers to improve the adhesion of mesenchymal stem cells (MSCs) [116]. Higher $CO_2$/$C_2H_4$ ratios yield well-defined actin microfilaments in MSCs, whereas lower ratios result in poor cell adhesion and survival. Moreover, other works show that $CO_2$-$C_2H_4$ plasma polymerization can result in COOH plasma polymer layers deposited onto polycaprolactone (PCL) nanofibers for diabetic wound healing [117].

- Fluorinated gases and vapors: fluorine ($F_2$) and carbon tetrafluoride ($CF_4$) plasma treatments can introduce fluorine-containing functional groups to improve the adhesion properties of polymer film such as polyethylene [118, 119], but can also etch and roughen the surface of polyamide [120]. Furthermore, $CF_4$ and hexafluoroacetone ($C_3F_6O$) are employed in the synthesis of fluorocarbon films with hydrophobic properties [121, 122].

The case of fluorinated gases and vapors deserves particular attention, especially the perfluorocarbon compounds ($CF_4$, $C_3F_8$, $C_{10}F_{18}$), owing to the complex dynamics they exhibit when introduced in cold plasma. These compounds create a scenario where two antagonist mechanisms—etching and deposition—coexist [123]. The predominance of one or the other mechanism mainly depends on the Fluorine/Carbon (F/C) ratio and the DC voltage applied to the substrate where the polymer film is synthesized. As Coburn et al. illustrate in **Figure 6**, when using a silicon substrate, only a $C_2F_4$ plasma (with F/C = 2) contributes to the deposition process, resulting in the formation of perfluorinated polymers (PTFE), as proposed in Equation (1). Conversely, a $CF_4$ plasma (with F/C = 4) is only involved in the etching process, leading to the formation of volatile species such as $SiF_4$, as detailed in Equation (2) [124]. This clear-cut division dissipates when cold plasma is supplied with $C_4F_{10}$ or $C_2F_6$, thereby invoking a competition between the two aforementioned mechanisms.

$$n\,CH_2F_2 \xrightarrow{Deposition} (CF_2)_\infty + Products \qquad (1)$$

$$SiO_2 + 2CH_2F_2 \xrightarrow{Etching} SiF_{4(g)} + Products \qquad (2)$$

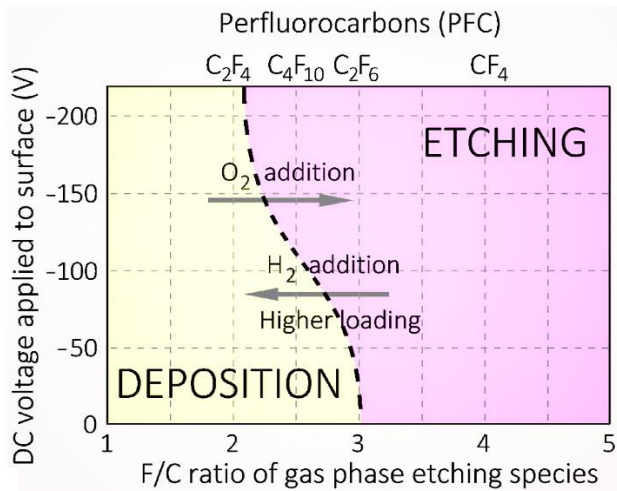

*Figure 6. Influence of the DC bias voltage and the F/C ratio (from fluorine and carbon atoms contained in the reactive gas) on the reaction behavior on the substrate. Addition of hydrogen or oxygen favors the formation or the etching of a polymeric film, respectively. Adapted from [129].*

A notable factor in this dynamic is the incorporation of a secondary gas, such as molecular oxygen or hydrogen, which can potentially bias the process in favor of one mechanism at the expense of the other. For instance, introducing oxygen can stimulate the formation of volatile etching species such as $COF_2$, CFO and $CO_2$, thereby fortifying the etching process. This effect is also observed when oxygen reacts with carbon in the plasma, leading to the generation of volatile CO and $CO_2$, again promoting the etching process. Contrastingly, the inclusion of molecular hydrogen results in the formation of new species such as HF, which absorbs F atoms (primary etchant species for silicon substrates). This process ultimately results in a decline in the F/C ratio, which in turn increases the deposition mechanism.

### 2.3.3. Chemical species densities in the gaseous phase

The densities of gaseous species play a critical role in defining the dynamics and characteristics of the plasma and, consequently, its applications. In the context of poly(2-vinylpyridine) (P2VP) exposed to a low-pressure RF plasma supplied with molecular oxygen, it appears that an increase in atomic oxygen density (O) accelerates the breakdown of polymer chains and, therefore, the polymer etching, while an increase in molecular oxygen ion density ($O_2^+$) escalates ion bombardment and sputtering, potentially contaminating the substrate [125]. The importance of oxygen density in etching processes is also evidenced when PMMA samples are exposed to He-$O_2$ plasma. Hence, for oxygen densities increasing from $7 \times 10^{13}$ cm$^{-3}$ to $1.1 \times 10^{14}$ cm$^{-3}$, the etching rate of PMMA increases from 47 µg·cm$^{-2}$·s$^{-1}$ to 67 µg·cm$^{-2}$·s$^{-1}$, also driving to different surface chemistries and roughening [126].

In addition to reactive species densities, monomer densities affect the properties of the plasma-synthesized polymer films. In the case of a low-pressure argon discharge supplied with a monomer vapor of trimethylsilyl acetylene (TMSAc), an increase in TMSAc density from 1.7 to 6.5 sccm results in the formation of $SiO_xH_yC_z$ films enriched with larger amounts of carbon, the promotion of Si-O bonds rather than Si-C bonds and WCA values increasing from 85.5° to 96.5° [127]. In a parallel example, increasing densities of allylamine monomer within a helium discharge modify the nitrogen incorporation in the resulting plasma-polymerized films [66]. As monomer gas flow rates drop from 1.0 slm to 0.2 slm, the N/C ratio rises from 0.21 to 0.24, while % C-NH$_x$ slightly increases (92.0 to 92.9) and % O=C-N decreases (from 8.0 to 7.1). This reflects an improved nitrogen incorporation in the polymer due to higher energy per monomer molecule.

### 2.3.4. Ion energy

The kinetic energy of ions, under the influence of a low-pressure plasma, can be significantly amplified — up to 100 eV — as the electric field in the sheath accelerates these ions, thereby increasing their flux towards the surface [98]. Conversely, increasing the pressure promotes ion-neutral collisions, thus reducing ion energy and, therefore, diminishing surface modification and film growth efficacy [93]. At atmospheric pressure, ions typically possess low thermal energies (<1 eV), although they can reach energies of tens of eV when the streamers approach the polymer surface [128], thereby influencing polymer processing and properties. As indicated in **Table 2**, ion energy influences several processing parameters, including deposition rate, etching rate, crosslinking, film density and film composition.

**Table 2. Comprehensive summary of ion energy effects on polymer film characteristics.**

| | |
|---|---|
| **Deposition rate** | High ion energy generally increases the deposition rate, as demonstrated in the synthesis of polyterpenol thin films using terpinen-4-ol and Melaleuca alternifolia oil [129]. An increase in mean ion energy from 3 eV to 17 eV augments mass deposition rates from 5.6 to 42.3 μg·m⁻²·s⁻¹. Conversely, a decrease in ion energy can hinder this rate, as adsorption and desorption processes become dominant factors. |
| **Etching rate** | Ion energy has a substantial impact on the etching rate of polymer films. This effect is evident in the case of photoresist coatings (consisting of a Novolak polymer with a photoactive compound) exposed to SF₆ plasma. As shown in **Figure 7a**, the etching rate increases from 200 nm/min to 600 nm/min when the ion energy is increased from 30 eV to 200 eV [130]. Similar trends are observed with PET films exposed to Ar-O₂ plasma, where higher ion energies not only improve etching, but result also in a heater load of the substrate [75]. |
| **Cross-linking** | Nanoscratching experiments on plasma-modified LDPE show that increased ion energy fluence stimulates chain crosslinking, enhancing surface shear resistance [131]. Ion energy fluences of 0, 70 and 630 kJ·m⁻² result in friction forces of 10, 30 and 100 μN, respectively. A similar trend is seen in PDMS during plasma immersion ion implantation, where elevated ion energies trigger a linear increase in wrinkle amplitude and wrinkle wavelength, two key indicators of crosslinking [132]. |
| **Film density** | In the DBD polymerization process, ion energy plays a key role in promoting surface interactions that contribute to film densification [133]. Specifically, when the ion energy is increased (by controlling the applied power from 30 W to 70 W), the synthesized hydrogenated amorphous carbon (a-C:H) films present densities rising from 1.1 g·cm⁻³ to 1.4 g·cm⁻³. |
| **Surface morphology** | Ion energy can influence the morphology or surface topography of polymer films such as PEEK, as ions with higher energy can affect the surface diffusion or mobility of polymer chains, leading to films with different surface textures [134]. |
| **Surface composition** | Plasma source ion implantation (PSII) is a technique where the control of ion energy is straightforward as it is directly correlated with the applied voltage. CF₄ plasma generated in a PSII device can change the surface composition of LDPE films to improve their hydrophobicity. As reported in **Figure 7b**, maximum WCA are obtained at ion energies of −1 kV, with values peaking at 122° and 113°, respectively, 1 day and 28 days after plasma treatment. These improvements can be attributed to the substitution of hydrocarbon and oxygen groups by fluorocarbon bonds (CF₂ and CF₃) [135]. However, when ion energies are further reduced to −5 kV and −10 kV, **Figure 7b** shows that the process becomes less effective, with WCA values close to 95°.<br>In plasma polymerization, ion energy has a significant impact on changes to surface composition, in particular on the effective integration of specific chemical functionalities. Analyzing the polymer coatings derived from an ethyl trimethylacetate (ETMA) monomer, Saboohi et al. identified two distinct scenarios: (i) under conditions of low ion energy and flux, ions gently interact with the surface, conserving the chemical integrity of groups and enhancing the overall film development; (ii) conversely, high-energy ions can cause the fragmentation of ETMA, a large molecular monomer, subsequently causing a deviation from the expected chemical functionalities [136]. |

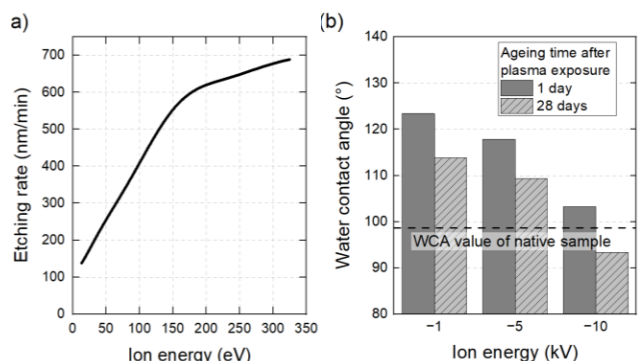

**Figure 7. (a) The etching rate as a function of ion energy of a crosslinked Novolak sample, thermostated at 20 °C during plasma treatment [130]. (b) Influence of ion energy from the plasma phase on LDPE surface wettability, 1 day and 28 days after plasma treatment [135].**

### 2.3.5. UV radiation

In cold atmospheric plasma, high-energy electrons interact with atoms and molecules, causing ionization and bringing electrons into excited states. When these excited electrons return to their ground state, they release their excess energy in the form of photons. If the photon energy is associated with a wavelength between 100 nm and 400 nm, it results in UV radiation emission. Between 10 nm and 200 nm, it results in VUV radiation that is so named because it is strongly absorbed by air and must, therefore, be studied in a vacuum environment. It overlaps with the extreme end of the UVC range and extends into even shorter wavelengths. Plasma UV radiation can result in the breakdown and crosslinking of carbon chains in the uppermost layers of the polymer, leading to enhancements in the polymer's durability and mechanical properties [137, 138, 139]. Experiments involving nanoscratching, as conducted by Tajima et al., demonstrate an appreciable increase in surface shear resistance of plasma-modified LDPE. This improvement is attributed to chain crosslinking stimulated by UV radiation [131]. On shorter wavelengths, VUV radiation can penetrate polymer films over several tens of nm (**Figure 3**), hence triggering stronger crosslinking, as evidenced by Narimisa et al. on polyolefins [140].

### 2.3.6. Plasma gas temperature

Gas temperature is a key determinant of reactive plasma chemistry, which in turn has an impact on the efficiency of surface activation processes. For example, in a cold plasma of ambient air, ozone production mainly occurs at 30 °C, as shown in **Figure 8a**. Conversely, when the temperature exceeds 200 °C, a significant production of nitrogen oxides (NO, NO₂) is observed, while ozone is no longer generated [141, 142]. Plasma gas temperature can therefore play a critical role in the nature of the gaseous reactive species generated (here, NOₓ vs. O₃) and subsequently affect the nature of the chemical groups functionalizing polymer surfaces (see **Section 3.2.6**).

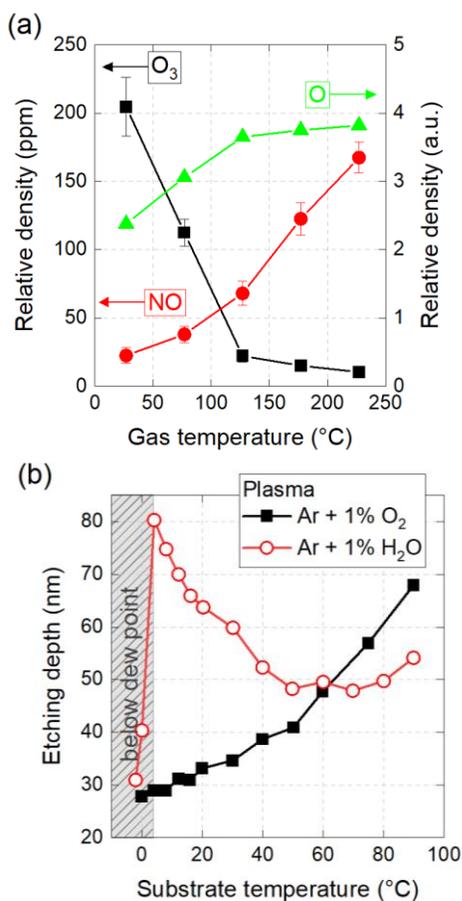

**Figure 8. (a)** Influence of plasma gas temperature on the relative density of reactive oxygen species measured either by optical emission spectroscopy (O) or gas tube detector (NO, O₃) [141]. **(b)** Influence of substrate temperature on the etching depth of PS films processed by plasmas of argon with 1% water (Ar + 1% H₂O) and argon with 1% oxygen (Ar + 1% O₂). Adapted from [112].

It is recommended to find an optimum balance in the plasma gas temperature, which must be high enough to promote surface activation processes, while remaining low enough to avoid polymer thermal degradation. For example, in HDPE surfaces, higher temperatures facilitate the selective abrasion of low-molecular-weight materials, thereby modifying surface topography [143]. In the case of PET, increasing the plasma gas temperature promotes the surface activation energy, although a threshold of 70 °C is recommended [144]. Taking this into account, wettability properties can be improved with WCA values falling from 65° to 45° only when increasing temperature from 35 °C to 70 °C. Finely controlling and reducing gas temperature can also be achieved by innovating specialized plasma sources such as liquid-film dielectric barrier devices (LF-DBD) [145]. Comparing LF-DBDs and conventional DBDs, which operate at 180 °C and 300 °C, respectively, under similar experimental conditions, PTFE samples exposed to LF-DVDs show less surface damage, higher concentrations of nitrogen- and oxygen-containing functional groups and WCA values as low as 65°.

### 2.3.7. Substrate temperature

Alongside the plasma gas temperature, substrate temperature (Tsubstrate) plays a pivotal role in altering the surface characteristics of polymers, especially in the etching/deposition rates of plasma-synthesized films. For example, Callahan (2001) reported a considerable increase in the etching rate of parylene-N in an oxygen plasma environment (400 mTorr, microwave plasma) from 5 nm/min to 70 nm/min as the temperature rises from 100 °C to 150 °C [146]. Interestingly, these thermal effects can vary and even reverse, depending on the nature of the plasma gas. This is shown in **Figure 8b** concerning the etching of PS films, where a rise in Tsubstrate from 10 °C to 100 °C increases the etching depth from 30 nm to 67 nm in Ar-O₂ plasma, while this parameter simultaneously decreases from 75 nm to 55 nm under Ar-H₂O plasma conditions [112].

Furthermore, Tsubstrate influences the dynamics of film deposition processes, as for 2-oxazoline-based polymer coatings deposited by an atmospheric pressure plasma jet (APPJ). In this case, an increase in Tsubstrate from 50 °C to 100 °C reduces the thickness from 84.8 nm to 57.0 nm due to the higher desorption rate, enhanced surface mobility and higher rates of the reactions consuming the depositing species on the substrate [147]. However, such a trend is not necessarily linear and largely depends on the plasma source itself. In the works of Mazankova et al., the influence of Tsubstrate on the thickness of plasma polyoxazoline thin films follows a bell-shape curve, initially rising from 600 μm (20 °C) to 2140 μm (120 °C) before falling back to 650 μm (150 °C) [148].

More generally, raising the substrate temperature can confer additional surface characteristics, as observed in the two previous studies (2-oxazoline-based polymer and polyoxazoline thin films). Surface techniques such as time-of-flight secondary ion mass spectrometry and XPS reveal that increasing Tsubstrate promotes crosslinking density within the film, which significantly improves their stability [147, 148]. Furthermore, higher values of Tsubstrate lead to lower surface oxidation of tetramethylsilane films [149] and terpenoid-derived plasma polymers [150]. A slight increase in surface roughness and reduction in wettability properties are also observed in that later case.

### 2.3.8. Plasma operation time vs. plasma exposure time

A clear distinction between plasma operation time and plasma exposure time (also called plasma treatment time) is essential. The issues behind these concepts can be understood considering a plasma source the ignition of which leads to a gradual heating of its components (electrodes, inner walls, dielectric barrier, sample, etc.). In the case of a plasma jet supplied in helium (6 slm, 10 kV, 10 kHz), **Figure 9a** shows a strong increase in the plasma source temperature (from 25 °C to more than 150 °C), followed by a thermal equilibrium at 180 °C after 45 min of operation. According to **Figure 9a**, treating PTFE samples for an exposure time of 3 min would lead to WCA values comprised between 104° (at t_operation = 0 min) and 122° (at t_operation = 45 min). The parameter implicitly incriminated here is the glass transition temperature of PTFE (about 120 °C), since all the other experimental conditions are unchanged [151]. The heterogeneity in these WCA values must therefore be attributed to the existence of a transient thermal regime, sometimes unbeknownst to the experimenter.

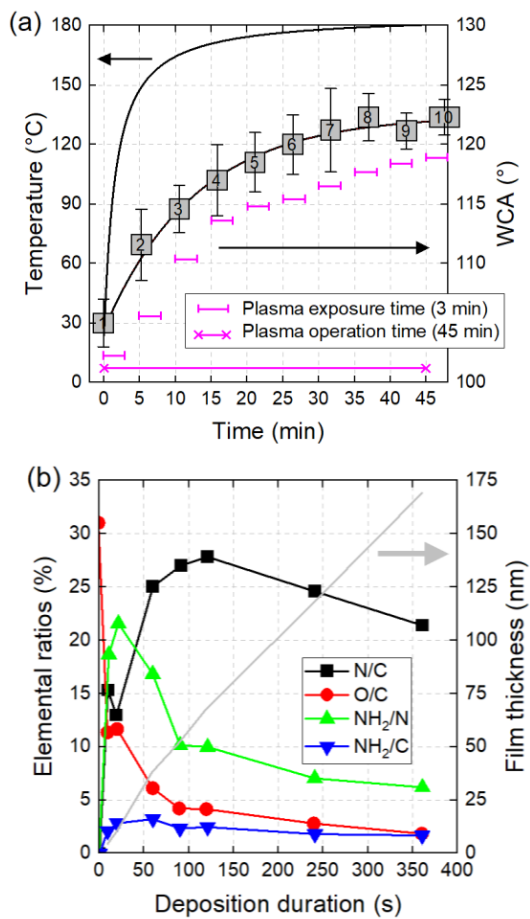

*Figure 9. (a) Influence of plasma operation time on the gas plasma temperature. Ten different PTFE samples are exposed for 3 min to plasma but at different instants and therefore different temperatures. Their WCA values are also reported. (b) Variation of the N/C, O/C, NH₂/C and NH₂/N ratios as a function of plasma deposition duration. Adapted from [152].*

With the clear distinction now drawn between plasma operation time and plasma exposure time, it is critical to underscore the non-linear effects that are exclusively tied to the plasma exposure time. Polymer film properties (texturing, functionalization, adhesion, etc.) can non-proportionally change relative to this characteristic duration. The reason is that different physico-chemical mechanisms may simultaneously occur with their own kinetics before converging towards an equilibrium (see CAP mechanism detailed in **Section 4.3**). This is illustrated in **Figure 9b**, where the deposition of cyclopropylamine-based plasma polymer films at $0.46 \text{ nm·s}^{-1}$ can result in a thickness linearly increasing with time, contrarily to its chemical surface composition monitored through the N/C, O/C, NH₂/C and NH₂/N ratios [152]. Vin Chan et al. explain the increasing-decreasing profile of the N/C ratio (peak of 28% at 120 s) resulting from the competition between etching and deposition processes. The O/C ratio drops from 31% to 2% due to the growth of film thickness, which lowers the number of oxygen photoelectrons coming from the underlying substrate. The amine grafting efficiency is assessed by the NH₂/C ratio, which increases from 0 to a maximum value of 3.2% (t = 60 s) and then slightly decreases due to longer exposure of the deposition film to UV radiation and active plasma species. Finally, the NH₂/N ratio, which stands for amine grafting selectivity, increases to 21.6% (t = 20 s) and then decreases to 6.2% (t = 360) due to a combination of etching and deposition caused by longer exposure to plasma. Other works underline non-linear variations of chemical surface composition vs. plasma exposure time, whether for the synthesis of PDMS-like coatings from hexamethyldisiloxane (HMDSO) monomer [153] or for the activation of PTFE surfaces [154].

# 3. Plasma modification of polymer surfaces

## 3.1. Positioning of plasma processes in relation to other approaches for modifying polymer surfaces

Enhancing the surface characteristics of polymers is essential to meet the various application requirements of industries such as electronics, biomedical, automotive and packaging. Several key techniques have proven effective in this domain:

- Wet chemical treatments involve soaking a polymer substrate into a chemical solution to modify its surface. An example of this process is acid etching (e.g., sulfuric, nitric or chromic) to enhance the surface roughness of polymer films, such as polyethylene [155]. Although simple, this technique demands strict control over reaction conditions to avoid over-etching, as well as adequate safety measures due to the use of potentially hazardous chemicals.

- Laser treatment can modify polymer surfaces by creating micro to nanostructures using adjustable parameters such as wavelength, pulse duration and fluence [156]. This process has found significant application in the medical sector, such as improving the shear bond strength between PEEK and resin-based luting agents to enhance the performance of fixed dental prostheses [157].

- UV/Ozone treatment is a non-thermal and dry processing method where UV light triggers the formation of ozone from ambient oxygen. This process is employed to oxidize and then enhance the wettability and adhesion strength of polymer surfaces (e.g., ethylene propylene diene methylene rubber, polyvinyl chloride and acrylonitrile butadiene styrene) [158].

- Flame treatment consists of exposing the polymer surface to a flame, typically generated by a gas burner. This heating source oxidizes the surface, improving its wettability and adhesion properties. The technique is commonly used in the packaging industry, for instance, in the treatment of bottle caps to improve the adherence of inks and labels [159]. Flame treatment also finds an interest in the automotive industry for pre-treating polymer components, such as polypropylene before painting or bonding [160].

- Mechanical abrasion techniques such as sandpaper or advanced sandblasting methods can be employed to

introduce roughness to polymer surfaces such as HDPE, LDPE, PP and silicone [161]. This roughness boosts the subsequent adhesion of coatings. Conversely, sand-in methods can also be utilized to achieve a durable and robust superhydrophobic surface with excellent water repellency and anti-icing properties [162].

While each technique offers distinct advantages and applications, plasma treatment stands out due to its unique characteristics. The plasma approach is indeed versatile and can uniformly modify complex shapes while preserving the bulk properties and improving adhesion without resorting to harsh chemicals or leaving residues. These aspects are further discussed in **Section 5.1.**

## 3.2. Surface properties modifiable by cold plasma

### 3.2.1. Surface etching

Etching refers to the process of removing material (e.g., low molecular weight fragments) from the surface of a polymer, while reducing, maintaining or increasing the surface roughness (See **Section 3.2.3**). With semi-crystalline PET polymers, for example, etching selectivity is naturally predominant, as amorphous regions are more sensitive to etching than crystalline ones [163]. In others, etching selectivity necessitates the use of masks, such as when PS-coated silica spheres are used to create a pillar array structure on PMMA plates (**Figure 10a,b**) [164]. The unique characteristics of plasma lead to the differentiation of two surface etching processes: (i) physical etching, related to the plasma sputtering process, and (ii) physico-chemical etching, correlated with the ion bombardment plasma process.

Strictly speaking, "chemical etching" as a stand-alone process is a misnomer because pure chemical reactions devoid of any physical influence are practically non-existent in a plasma environment. Plasma introduces a plethora of energetic ions, electrons and excited species, all of which can physically bombard or interact with a polymer surface. Although chemical reactions occur when reactive plasma species interact with the polymer surface to produce volatile by-products, these reactions are usually associated with simultaneous physical bombardment by energetic plasma entities.

Physical etching occurs when high-energy particles, such as ions or neutral atoms, interact with a polymer surface, transferring their kinetic energy to physically dislodge atoms from the polymer. This requires ion density to be very high and ion kinetic energy to exceed the bond dissociation energy (BDE) of the surface atoms, as reported in **Table 3**. While the process is highly effective at ion energies above several hundred eV for low-pressure plasmas [165], it can severely damage polymer surfaces due to internal cascade collisions, resulting in almost no selectivity. Such physical etching can be performed by low-pressure RF plasma sources supplied in argon to create super-hydrophobic PP or PTFE surfaces, with WCA values of approximately 170° [166]. Similar results are obtained on PTFE samples exposed to an atmospheric RF plasma torch for which XPS analysis indicates the absence of surface oxidation, while significant masse losses evidence a physical etching [167].

**Table 3.** *Typical bond dissociation energies (BDE) encountered in polymers* [168].

| Bond | BDE (eV) | Bond | BDE (eV) |
|------|----------|------|----------|
| H–H | 4.36 | H–C | 4.13 |
| C–C | 3.48 | H–N | 3.91 |
| N–N | 1.70 | H–O | 3.66 |
| O–O | 1.45 | C = C | 6.14 |
| C–I | 2.16 | O = O | 4.98 |
| C–N | 3.08 | C–O | 3.60 |

The physico-chemical etching of a polymer is initiated by the bombardment of the surface by positive and/or negative ions, as well as neutral reactive species, all generated by the cold plasma. For example, oxygen radicals (O•) from $O_2$ microwave plasmas interact with polymers, leading to the emergence of volatile oxides. At the same time, fluorinated entities (F–, F+) from $SF_6$ microwave plasmas serve as predominant etching agents [169]. This bombardment leads to the ejection of atoms from polymer surface layers, producing nanostructures such as pits and pores. In addition, it can break chemical bonds, generating new reactive species that amplify the physical etching process. This phenomenon is documented by Fricke et al. when studying the etching of PEEK by an Ar-$O_2$ plasma jet [170]. Among the various physicochemical processes, reactive ion etching (RIE) stands out. In RIE, a low-pressure RF plasma, typically CCP or ICP, achieves etch rates ranging from a few nanometers to several hundred nanometers per minute. Under ion bombardment etching conditions, Y. Ohnishi et al. demonstrated that the etching rate of polymers is primarily determined by their "effective" carbon content, which is calculated by subtracting the number of oxygen atoms from the carbon atoms in the polymer [171]. This distinction suggests that C-O or C=O groups in polymers have a higher sputtering yield than pure carbon. Notably, the etching rate linearly depends on the N/(NC−NO) ratio, where NN, NC and NO are the total number of atoms, carbon atoms and oxygen atoms, respectively. All components of the -CHx- type in polymers, where x varies from 1 to 3, exhibit identical sputtering yields. Interestingly, etch resistance is almost the same for polymers solely composed of C, O and H atoms as for those containing Cl, F or N atoms. Chemical bond strength does not appear to play an important role in etch resistance, mainly because the energy of the incident ions during etching exceeds the energy of the chemical bonds, particularly in ion-beam etching scenarios [171].

As part of RIE, several issues can adversely affect the etching rate and aspect ratio of the cavities, including:
- The micro-loading effect (or lag effect) illustrated in **Figure 10c** refers to a decline in the etching rate observed in areas with densely packed features. The decrease is primarily due to a competition between these closely situated features to interact with a limited number of reactive species from plasma. This competition, combined with the inefficient removal of by-products, leads to non-uniformities across the substrate, as sketched in **Figure 10c** [172].
- Aspect Ratio Dependent Etching (ARDE) is a phenomenon in which the etching speed varies according to the aspect ratio of the features undergoing etching. As shown in **Figure 10d**, a high aspect ratio correlates with a reduced etching speed. **Figure 10d** shows that an increase in the aspect ratio drives a decrease in the etching rate due to the difficulty of reactive ions reaching the bottom of the feature and the difficulty of etch by-products diffusing out. This can lead to non-uniform etch profiles and "bowing" or "notching" in the etched sidewalls [173]. Conversely, an inverse ARDE effect can also be observed depending on the specific materials being etched, the plasma chemistry, the design of the RIE system and the process parameters. In this case, the etch rate in a high-aspect-ratio feature is faster than that in an open area. This can be attributed to various factors: differences in passivation layer formation [174], charge accumulation at the bottom of high-aspect-ratio features thus locally enhancing the electric field [175] or ion focusing/funneling into narrow features, thus leading to a higher ion concentration [176].

- Sidewall bowing in plasma reactive ion etching (RIE) corresponds to a non-vertical etching profile where the sidewalls of the etched feature curve inward or outward, creating a 'bow' shape (**Figure 10e**). This effect typically arises from variations in the etch rate over the feature's depth, leading to dimensional inaccuracies or structural instability in microfabricated structures [177]. Another possible anomaly is notching, as represented in **Figure 10f**.

- Undercutting refers to a phenomenon in which the lower parts of a patterned polymer are etched more than the upper parts, resulting in an undercut profile, which is represented in **Figure 10g**. While the ions primarily bombard the substrate from above, the plasma's chemical etching components can horizontally attack the polymer, below the protective mask [178]. Undercutting is specifically lateral etching under the mask, whereas overcutting refers to etching beyond the intended area or depth in any direction (**Figure 10h**).

- Chemical residues are non-volatile by-products resulting from interactions between plasma reactive species and the polymer to be etched. These residues can modify surface properties or form an undesirable barrier that prevents further etching [179].

- Mask scattering is when the high-energy ion bombardment physically sputters or chemically erodes the etching mask before scattering it across the substrate. Then, the scattered mask particles can settle on the substrate, causing irregular etching patterns, since they effectively act as a secondary, unintended mask [180].

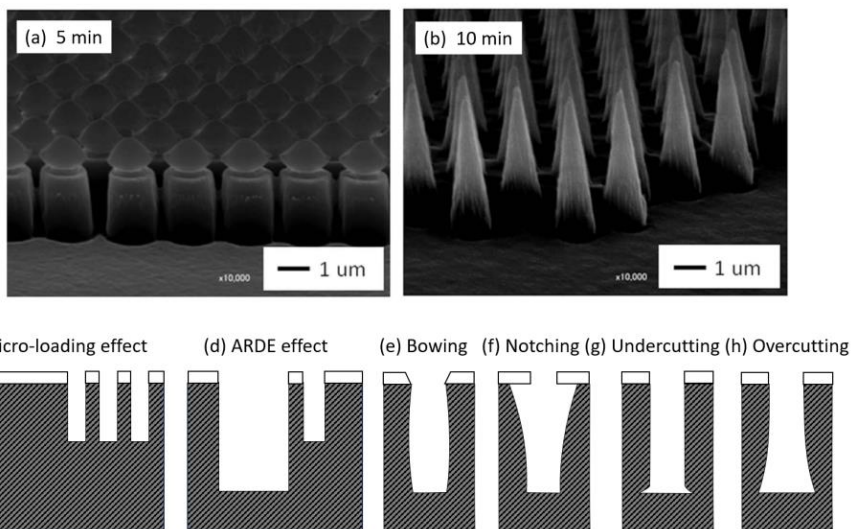

**Figure 10. (a,b) SEM photographs showing pillar and cone structures created on PMMA sheets. These structures are formed using 2.1 µm diameter PS spheres as etching masks (reproduced with permission) [164]. (c–h) Schematics of the main anomalies encountered in plasma RIE [180].**

Several strategies can limit these anomalies, including pulsed plasmas, UV radiation, cooling mechanisms, magnetic-field-induced beam control and lithography [181, 182, 183, 184]. In addition, the precise manipulation of parameters such as gas composition, pressure and power enables fine-tuning of the etch rate, directionality and process selectivity, helping to reduce anomalies [185]. The resulting versatility and precision explain the success of RIE in the semiconductor industry, although the ever-scaling down of semiconductor devices (memory devices, computing elements, etc.) highlights the emergence of new constraints, including etching profile control, short channel effect and material selection [186]. Today, a more refined approach is offered by atomic layer etching (ALE) to remove material layer by layer with atomic precision. This involves a self-limiting, cyclic process utilizing selective chemical reactions and passivation steps, allowing for the precise control and preservation of underlying layers [186].

### 3.2.2. Surface cleaning

Cold plasma processes can be developed either to clean various substrates from polymer residues or, conversely, to clean polymer surfaces from contaminants. In any case, plasma cleaning is specifically designed to preserve the inherent properties of the surface so as to not induce significant surface alterations.

Cleaning substrates from polymer residues is appropriate during the synthesis of materials such as graphene, where unexpected PMMA residues are produced [187]. Their removal from a single graphene layer can be achieved by $H_2$ or $H_2-N_2$ plasma without causing any damage, thus making the process both efficient and selective. Another example is the fluorocarbon plasma etching of low-k material such as porous dielectric SiCOH [188]. While $CF_X$ residues are formed on the sidewalls, they can be removed by high-energy H atoms (exceeding 1 eV) generated in a $He/H_2$ plasma, without harming the low-k material.

Cleaning polymer surfaces from contaminants means removing organic residues, oxides and other undesirable substances such as air pollutants and fingerprints [189]. The plasma often contains reactive species such as ions, electrons and radicals, which interact with the surface and remove contaminants. Furthermore, plasma cleaning can generate chemically active sites on the polymer surface that are beneficial for improving wettability or adhesion.

When it comes to eliminating biological agents such as fungi and bacteria, the terminology of plasma decontamination is preferred to that of plasma cleaning. Plasma's reactive components can cause cell wall rupture, (plasmid) DNA oxidation and other damaging effects, making it particularly useful in fields such as food processing and healthcare [190, 191]. When the aim is to thoroughly clean a polymer surface by eliminating all micro-organisms, this is generally referred to as plasma sterilization. Low-pressure SF6 plasma is particularly effective for sterilizing polymer surfaces such as PE, PET and PVC. This is demonstrated by the survival rates of various bacterial strains (including *Streptococcus spp., B. cereus, Proteus spp., S. aureus and E. coli*, among others), which show a reduction in colony-forming units (CFU) of over nine decades when exposed to treatment times of at least 3 min [192]. In medical or pharmaceutical contexts where sterility is a prerequisite, plasma sterilization offers advantages over conventional techniques (autoclaving, chemical sterilants), as it enables heat-sensitive materials to be sterilized, and complex shapes or tight spaces to be accessed [193].

### 3.2.3. Surface Roughening and Surface Texturing

Cold plasma can significantly modify the topography, roughness and texture of a polymer surface. While these terms are often interchanged as synonyms, it is important to stress their nuances:

- Topography vs. roughness: The key difference here is the scale of the features. Topography includes all the features on the surface at various scales, while roughness is specifically concerned with the fine-scale deviations from an ideal smooth surface. As an example, the AFM pictures in **Figure 11a-d** indicate an increase in the surface roughness of PTFE samples after different exposure times to an RF plasma torch [194].
- Roughness vs. texture: Roughness is about the degree of deviation from a smooth surface without considering any directional pattern or regularity. Texture, on the other hand, implies a certain regularity or pattern in the arrangement of surface features.
- Topography vs. texture: Topography is a more general term that includes all surface features, while texture refers to the specific patterns or directional characteristics of these features. **Figure 11e-g** corresponds to AFM or SEM pictures showing drop-like, ripple and honeycomb patterns, respectively [195, 196, 197].

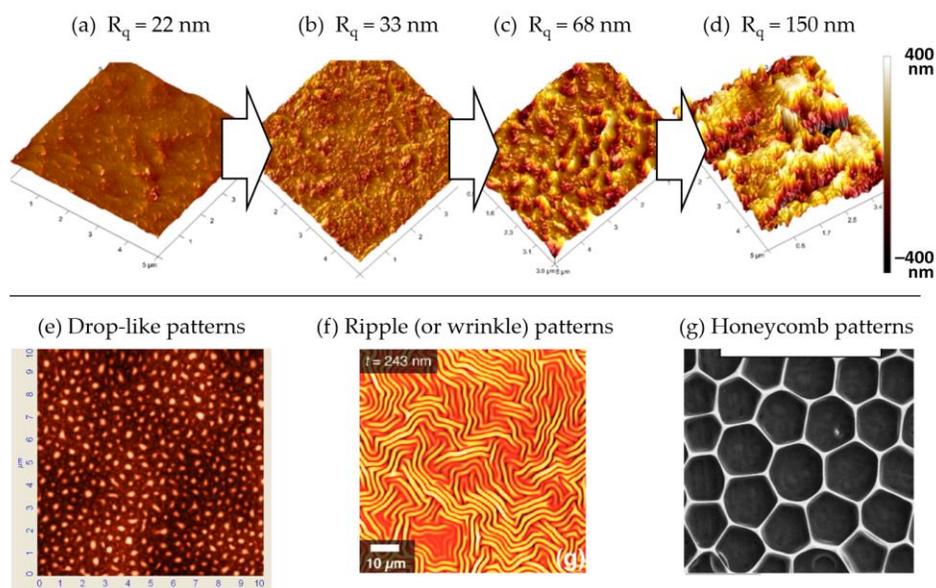

*Figure 11. (a–e) AFM pictures of PTFE surfaces exposed to an RF plasma torch for different exposure times; (a) 0 min, (b) 2 min, (c) 26 min, (d) 50 min. Rq is the root mean square roughness [194]. (e) AFM picture showing a drop-like pattern of plasma-treated PP [195]. (f) AFM picture showing ripple pattern of plasma-treated PDMS [196]. (g) SEM picture showing honeycomb pattern of plasma-treated FEP [197]. All images are reproduced with permission. Copyright (c) 2012 Wiley-VCH Verlag GmbH & Co. KGaA, Weinheim, Copyright (c) 2009, American Chemical Society, Copyright (c) 2019 Wiley-VCH GmbH.*

Increasing surface roughness amplifies surface area and reactivity, which is beneficial for many applications, including adhesion, wettability and bonding capacity for further processing [194]. However, these effects depend on the specifics of the roughening process and the requirements of the application. Consequently, surface roughness not only diversifies surface texture, but also has an impact on the polymer's physical and chemical properties. Complementary information is given in **Section 3.2.8** considering the importance of nanoscale and micrometer-scale roughness in the context of surface wettability issues.

### 3.2.4. Surface crosslinking

High-energy ions from the plasma can break bonds within polymer chains at the surface, hence leading to the creation of unpaired electron bonds that are left "dangling" due to surface effects [198].

These resulting dangling bonds may then interact with each other or with additional plasma species, thus triggering crosslinking between the polymer chains. While crosslinking is generally less prevalent in polymers prepared following conventional protocols, it is substantially higher in plasma-modified or plasma-synthesized polymer films, as represented in **Figure 12a**, where crosslinking forms a three-dimensional network structure. This interconnection of polymer chains can be achieved through the formation of covalent bonds, ionic interactions or other types of chemical bonds between the chains. Crosslinking can result in significant changes in the mechanical, thermal and chemical properties of the polymer surface, often leading to enhanced strength, durability and resistance to solvents, heat or chemical degradation [199].

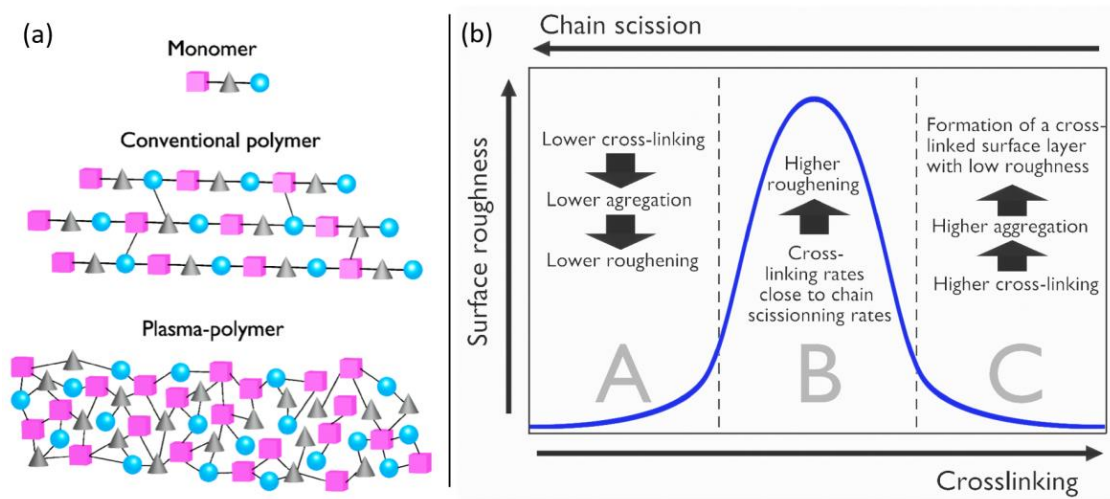

*Figure 12. (a) Diagram illustrating the structural differences between a basic monomer molecule, a conventional polymer and a plasma-polymer. A polymer is a large molecule composed of repeating units called monomers that bond together in a chain-like structure. The molecular weight of a polymer reflects the number and type of monomers used and typically varies from 1000 to several million Daltons (or amu) [200]. (b) Model explaining how surface roughness (blue curve) is impacted by crosslinking and chain scission. Both mechanisms can aggregate surface polymer patterns, likely to induce surface roughness during the plasma etching process. Adapted from [201].*

Crosslinking and chain scission are two simultaneous phenomena that play a crucial role in the surface roughening process during plasma etching. According to the model of Y.-H. Ting et al., chain scission reduces molecular weight, thereby increasing mobility, while the interconnection of adjacent polymer chains acts in the opposite direction via crosslinking [201]. This mechanism is visualized in **Figure 12b**, where surface roughness is distributed over three distinct regions, each correlating with the degree of crosslinking. In region A, a lower degree of crosslinking corresponds to minimal aggregation, resulting in negligible surface roughness. Region B shows an equilibrium between crosslinking and chain scission rates, triggering aggregation and subsequent surface roughness. In contrast, region C undergoes intensive crosslinking, forming a complete crosslinked layer that inhibits both mobility and aggregation, resulting in a smooth surface.

The balance between crosslinking and chain scission rates depends on both the nature of the polymers and the plasma conditions. Consequently, different polymers exposed to identical plasma conditions may exhibit different surface roughness. This is verified when ion bombardment energy enhances crosslinking in PS while not in PMMA, and poly(2-ethyl-2-oxazoline) where depolymerization dominates [198, 201].

### 3.2.5. Surface crystallinity

The crystallinity of a polymer film can be assessed by differential scanning calorimetry (DSC). This technique measures the amount of energy absorbed or released by the film as it is heated or cooled, unrevealing its glass transition temperature ($T_g$), melting temperature ($T_m$) and heat of fusion ($\Delta H_f$) [202]. This information is instrumental in distinguishing between crystalline and amorphous polymers: (i) a crystalline polymer has a specific melting point, while an amorphous polymer does not, (ii) the heat of fusion is related to the degree of crystallinity: the higher the heat of fusion, the higher the crystallinity of the polymer. Complementarily, X-ray Diffraction (XRD) can be achieved to determine the atomic and molecular structure of the (semi)crystalline polymer film [202, 203]. Crystalline polymers will give sharp and distinct peaks on an XRD diffractogram, indicating a regular arrangement of atoms. In contrast, amorphous polymers will provide broad peaks or a hump, indicating a more random arrangement of atoms.

As polymers solidify from a molten or soluble state, their chains can either form a crystalline (i.e., ordered) structure or an amorphous (i.e., disordered) one. Some polymers entirely lean towards one of these two states, while many others adopt a blend of both, becoming semi-crystalline [204]. In these, the crystalline regions are highly structured and compact, with polymer chains organized in a regular pattern, contrasting with the amorphous regions where chains are randomly intertwined, not oriented and exhibit chain mobility. This concept can be further explored considering the same RF oxygen plasma to treat semi-crystalline or amorphous PET surfaces. In comparison with the plasma-treated amorphous surfaces, the plasma-treated semi-crystalline exhibits the following features:

- Lower etching rates, leading to smaller weight losses [205]
- Increased surface roughness and higher incorporation of oxygen-based functionalities, as particularly illustrated in **Figure 13a** for PET surfaces exposed to an inductively coupled RF oxygen plasma. Consequently, semi-crystalline PET surfaces show higher levels of C-O, C=O and O-C=O functional groups, which contribute to superior wettability than amorphous surfaces [206]
- Enhanced thermal resilience: unlike the amorphous samples that heat up in 30 s, they can endure up to 2 min of plasma exposure without exhibiting thermal-induced damages [206]
- Slower ageing process, typically due to the restricted mobility of functional groups on the crystalline parts of the polymer [205]

The influence of surface crystallinity on surface aging deserves special attention, particularly when considering the recovery of hydrophobicity in plasma-treated polymers. Still considering PET surfaces exposed to an oxygen plasma, Hyun et al. demonstrated that hydrophobic recovery is intrinsically linked to the reduction of polar groups when the surface is exposed to nonpolar environments [207]. This stable fraction of polar groups can serve as a metric of a polymer's crystallinity. The distinction between the untreated surface (WCA = 75°) and a surface that has partly recovered hydrophobicity reveals the number of stable polar groups remaining in the polymer surface's crystalline region. As **Figure 13b** depicts, PET surfaces exhibit a WCA value close to 20° due to the addition of polar groups to the surface. However, after 50 days in air, these PET surfaces gradually regain their hydrophobic properties. For plasma-treated PET surfaces with higher crystallinity levels (as depicted by the blue curve), this process is somewhat limited, with WCA increasing from 15° to 42°. On the other hand, PET surfaces with lower levels of crystallinity (indicated by the red curve) exhibit a more robust recovery of hydrophobicity, with WCA reaching as high as 64°.

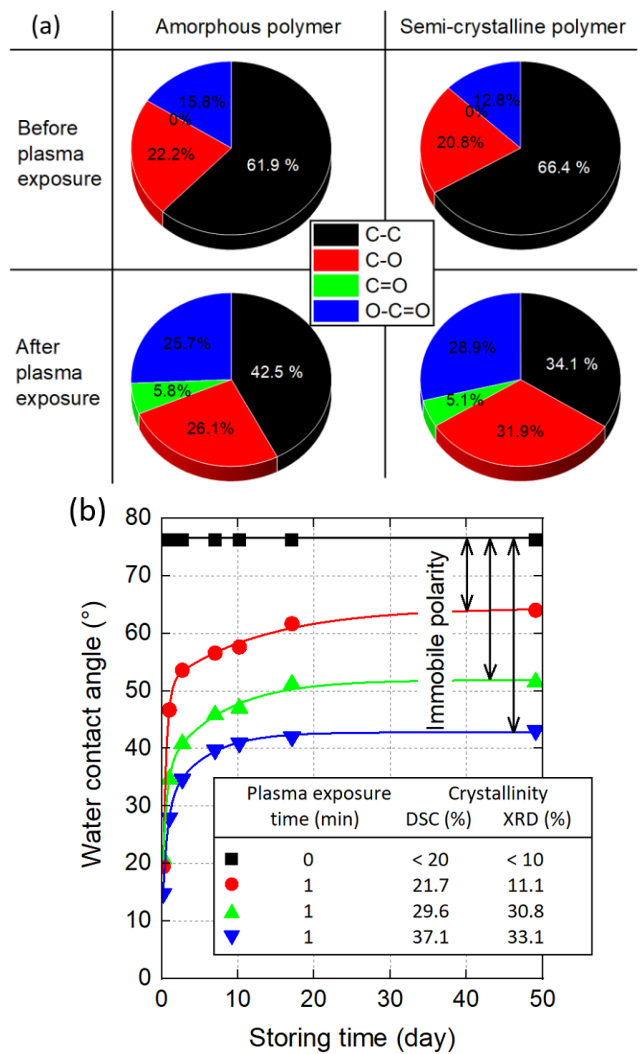

**Figure 13.** (a) Functional groups from C 1s spectra versus treatment time for amorphous and semi-crystalline PET foil (C-C at 285.0 eV, C-O at 286.5 eV, C=O at 287.5 eV and O-C=O at 289.0 eV). Adapted from [206]. (b) Effect of crystallinity on the surface dynamics ox oxygen plasma treated PET surfaces. Adapted from [207].

### 3.2.6. Surface functionalization by chemical activation

Chemical activation here refers to the chemical modification of a polymer surface, typically to enhance its wettability or adhesion properties, with spinoffs in biology such as improving cyto-compatibility, biocompatibility and bioactivity of polymer implants for drug delivery, cardiac tissue engineering, prosthetics and orthopedics [208, 209]. Plasma appears as a relevant approach for increasing polymer surface energy owing to its active species that can introduce various reactive functional groups. **Table 4** succinctly illustrates these functional groups depending on the gas source used and how these changes impact the polymer's properties. This knowledge is crucial for the tailored design and customization of polymer materials in various fields, including biomedical applications, material science and electronics.

*Table 4. Bridging the gaseous reactive species from plasma with the functional groups introduced on polymer surfaces.*

| Gaseous Reactive Species | Gas/Vapor Sources | Functional Groups Introduced on Polymer Surface | Effects & Applications |
|---|---|---|---|
| **Oxygen radicals** | $O_2$, water vapor | Carbonyl groups (C=O) Carboxyl groups (-COOH) Hydroxyl groups (-OH) [211] | Increase in the polymer surface energy that can then improve wettability, adhesion and compatibility with other materials [212]. |
| **Excited water species** | Water vapor | Hydroxyl groups (-OH) | Improving the wettability and adhesion of the polymer [213,214]. |
| **Nitrogen radicals** | $N_2$, ammonia ($NH_3$) | Amine groups (-NH₂) Amide groups (-CONH₂) Nitrile groups (-CN) [215,216] | Improving cell adhesion of PLA samples, as evidenced by MTT and SEM [217]. Enhancing reactivity and hydrogen bonding capabilities of polymer surfaces. |
| **Nitric oxide radicals** | NO, $NO_2$ | Nitro groups (-NO₂) Nitrite groups (-ONO) Amine groups (-NH₂) Amide groups (-CONH₂) | Polymer surfaces with antibacterial properties for biomedical applications [218,219]. |
| **Hydrogen radicals** | $H_2$ | Hydrogen-containing functional groups that are both stable and unreactive. | Plasmas can effectively clean surface contaminants (e.g., residual monomers, surfactants) from polyethylene, polypropylene and polyimide [220,221]. |
| **Carbon radicals** | $CH_4$, $C_2H_6$ | Methyl groups Ethyl groups Etc. [222,223] | The functional groups can modify the surface properties of the polymer, such as its hydrophobicity or conductivity [224]. |
| **Reactive oganic species** | Styrene, vinyltrimethylsilane (VTMS) or divinylbenzene (DVB) | Vinyl groups (-CH=CH₂) | Introducing unsaturation into a surface to make it more reactive and capable of undergoing further polymerization or crosslinking. |
| **Fluoro (carbon) radicals** | $SF_6$, $CF_4$ or fluorocarbon precursors. | Fluorine-containing groups ($-CF_x$) | Enhancing hydrophobicity, chemical resistance and non-stick properties. |

As previously mentioned, exposing a polyethylene surface to an oxygen plasma results in its activation, i.e., the introduction of polar functional groups, such as hydroxyl and carbonyl groups, which then increase its surface energy, thereby enhancing wettability. To decipher the interplay between a water droplet and this plasma-activated surface, it is essential to thoroughly examine this phenomenon as follows:

- First, water molecules in a droplet are bonded together through a combination of covalent and hydrogen bonds [224]. Covalent bonds, which are exceptionally strong due to shared electrons, unite the two hydrogen atoms with the oxygen atom within a single water molecule (**Figure 14a**). Conversely, hydrogen bonds link separate water molecules within the droplet. Here, the slightly positive hydrogen atom of one water molecule is attracted to the slightly negative oxygen atom of another water molecule (dashed lines in **Figure 14a**). While individually weaker than covalent bonds, the combined force of numerous hydrogen bonds imparts unique characteristics to water, including its high surface tension, heat capacity and ability to dissolve many substances.

- The outermost layer of polyethylene consists of carbon and hydrogen atoms (see **Figure 14b**). The electronegativity

values of these elements (2.55 for carbon and 2.20 for hydrogen) are so similar that the resulting C-H covalent bond can be considered non-polar. This means that the surface energy of polyethylene is extremely low and that no complete hydrogen bond can be formed between the H atoms from the PE surface and the O atom from the water droplet [225]. However, a native polyethylene surface is always, albeit minimally, oxidized, meaning that some sites present O atoms inserted between C and H atoms. With an electronegativity value of 3.44 for O, the C-H and C-O bonds are therefore covalently polar. The singular C-O-H site, as depicted in **Figure 14b**, can then create an effective hydrogen bond with the water droplet.

- **Figure 14c** depicts the ideal situation of a thoroughly oxidized PE surface following plasma exposure. As the topmost layer entirely consists of polar covalent bonds, the surface energy is high enough to surpass the water droplet's surface tension, causing it to reshape and maximize the droplet-PE interface area.

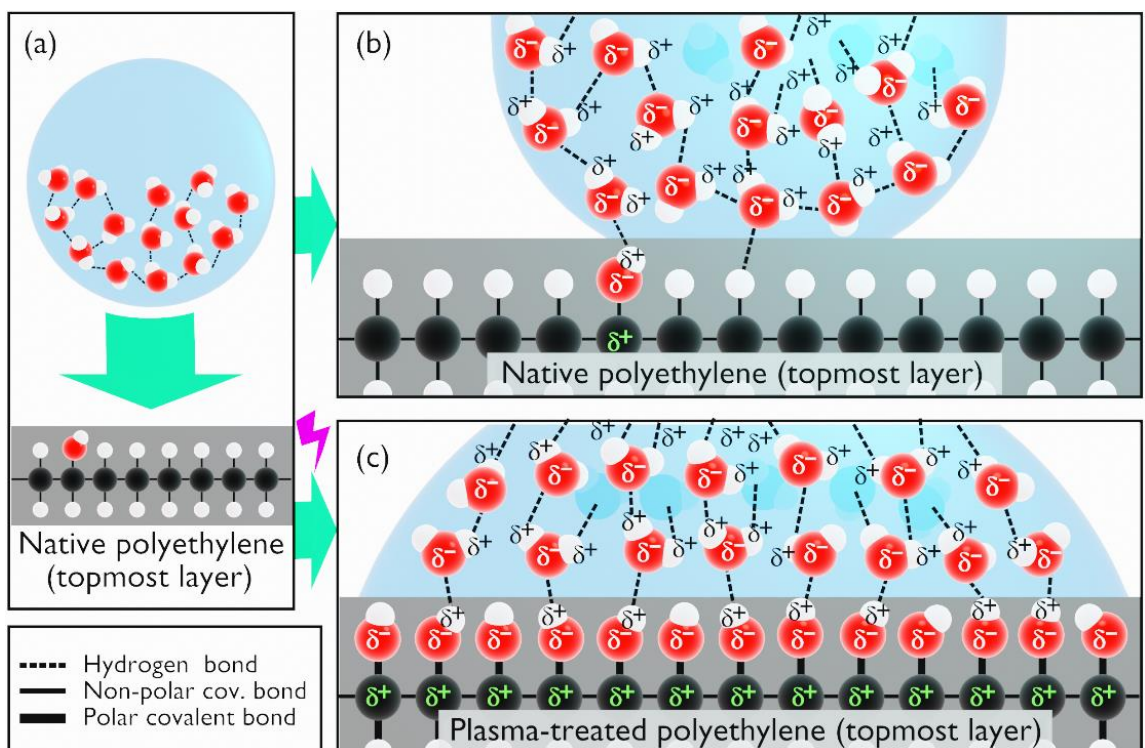

***Figure 14.** Schematics of (a) the topmost layer of a slightly oxidized polyethylene film before its contact with a water drop, (b) water drop in contact with the native polyethylene surface, showing hydrogen bonds within the drop and non-polar covalent bonds within the polymer, (c) water drop in contact with a plasma-treated polyethylene surface showing strong oxidation and subsequent polar covalent bonds within the polymer.*

The introduction of hydroxyl, carbonyl or carboxyl groups to the surface does not add new chemical elements, which allows this functionalization to be considered as a chemical surface activation. Similarly, the functionalization of a polymer surface by -NH$_X$ groups (or -CF$_X$ groups) while natively containing nitrogen atoms (or fluorine atoms) corresponds to a surface activation. However, strictly speaking, the introduction of -NH$_2$ groups onto a polyethylene surface is no longer plasma activation but plasma grafting, since new chemical elements, i.e., different from the native polymer surface, are introduced (see **Section 3.2.7**).

### 3.2.7. Surface functionalization by chemical grafting

Chemical grafting is the process of bonding new chemical species or carbon chains to the surface of a polymer, in order to impart additional characteristics such as biocompatibility or antimicrobial behavior [226, 227]. In plasma processing, this grafting process is facilitated by the introduction of a reactive gas or vapor that carries the desired chemical species into the plasma. These reactive species react with the functional groups present on the surface (formed during the surface activation process), establishing a covalent bond between the grafted species and the surface.

The anti-fouling properties of polyethersulfone (PES) ultrafiltration membranes can be improved by using a corona air plasma to graft polymer chains of hyperbranched polyethylene glycol (HB-PEG), following the mechanisms in **Figure 15a** [228]. Upon approaching the polymer surface, the oxygen atom from HB-PEG interacts with a sulfur atom from the PES membrane. There, the electron pairs are transferred to the S=O bond and then to the oxygen atom (which becomes negatively charged) while the oxygen atom of the OH functional group acquires a positive charge. This electron rearrangement forms an intermediate compound on which the hydroxyl group of a HB-PEG molecule bonds while a water molecule is simultaneously expelled.

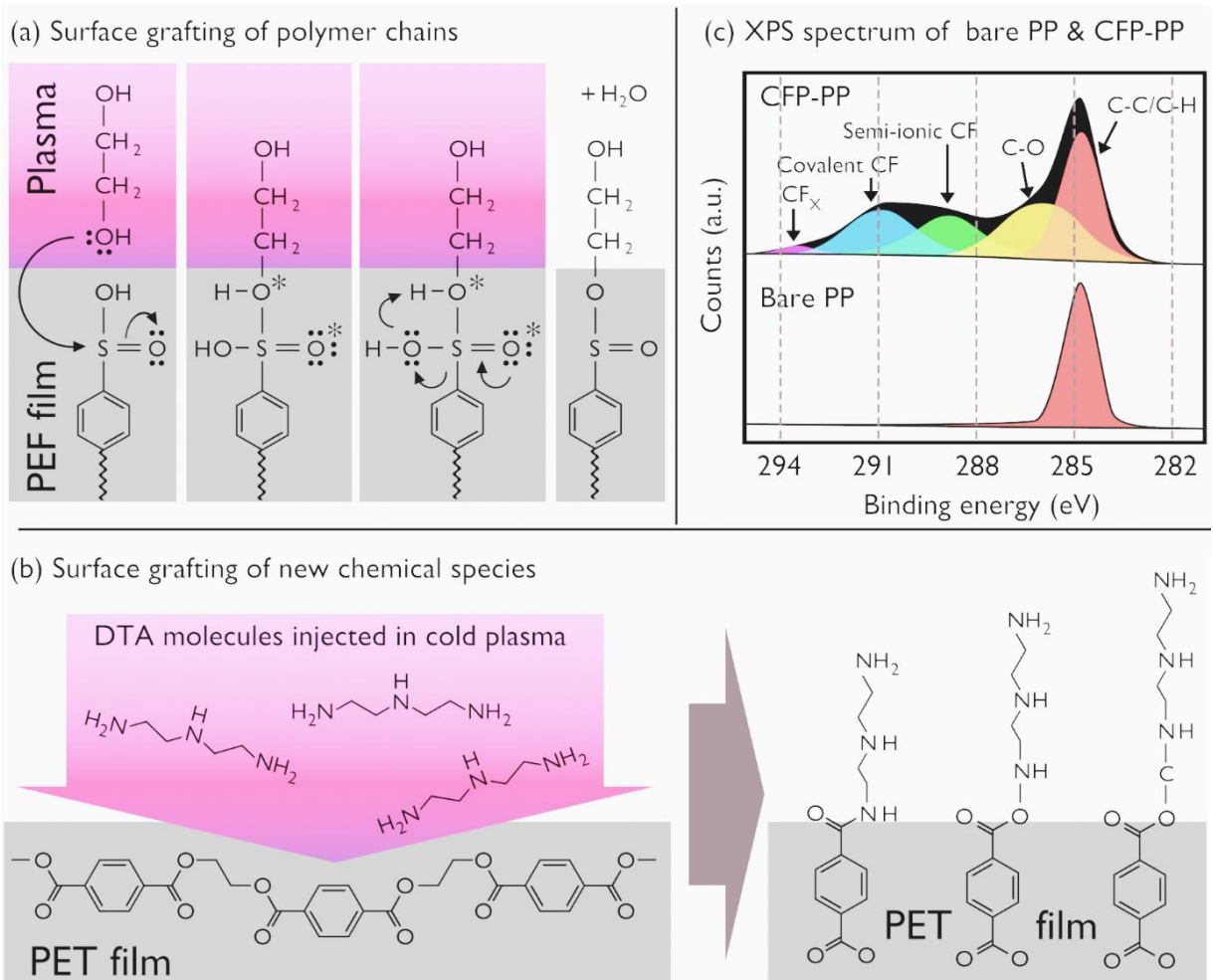

*Figure 15. (a) The reaction between the HB-PEG and surface-activated PES. Adapted from [228]. (b) Proposed scheme of some plasma treatment routes for PET waste films with DTA. Adapted from [229]. (c) High-resolution XPS spectra of C 1s. There are only signals for C-C and C-H bonds for bare PP. For CFP-PP, additional binding energies located at 285.9, 288.9, 290.9 and 293.3 eV are assigned to the C-O bonds, semi-ionic C-F bonds, covalent C-F bonds and CF₂ groups. Adapted from [230].*

Chemical grafting also extends to the covalent bonding of new chemical entities to a polymer surface. To enhance biocompatibility, Mora-Cortes et al. engineered a low-pressure RF plasma technique that facilitates the grafting of primary amine (-NH₂) groups onto a PET film using diethylenetriamine (DTA) [229]. Although the underlying mechanisms have yet to be completely unraveled, XPS analyses provide evidence that DTA grafting predominantly occurs at the O=C-O-C bond sites of the PET film. Additionally, the grafted functionalities exhibit variable oxygen and carbon atom counts, as depicted in **Figure 15b**. Shifting the focus from biocompatibility enhancement to battery technology optimization, fluorine-containing functional groups also play a crucial role in the chemical grafting process (**Figure 15c**). For improving lithium metal anode batteries, fluorine-containing functional groups are introduced onto a polypropylene (PP) separator with an RF CF₄ plasma treatment [230]. The resulting grafted polar groups enhance the PP separator's affinity for Li-ions, leading to improved wettability, ion conductivity and lithium-ion transference number. Furthermore, they also contribute to the formation of a LiF-rich solid electrolyte interface (SEI) film. The XPS tests in **Figure 15c** confirm the presence of fluorine-containing functional groups, with binding energies for C-O, semi-ionic C-F, covalent C-F bonds and CF₂ groups observed in the CFP-PP separator [230].

In addition to the grafting of amine groups and fluorine-containing functional groups, plasma processes can be employed for the grafting of nanoparticles [231]. Furthermore, 3D-printed fractal polymeric substrates can be plasma-activated to subsequently graft metal oxide nanoparticles (ZnO and TiO₂) and iron-based metal-organic framework (Fe-MOF) nanoparticles. Plasma is employed to create a strong and permanent bond between the nanoparticles and the polymeric surface. The resulting hybrid nanomaterials show potential applications in supported catalysis, such as the photocatalytic degradation of organic pollutants in water.

### 3.2.8. Surface wettability: a property resulting from chemical activation and roughening

While some plasma processes are designed to modify a single surface property, others are intentionally developed to induce both physical and chemical surface modifications. This is the case for surface chemical activation and physical surface roughening that, combined, modify the surface wettability. The Wenzel and Cassie-Baxter models are classic theories used to describe the wettability of a surface, especially in the presence of roughness or textures [232, 233]. These models, summarized in **Table 5**, are often invoked to explain the hydrophilic and hydrophobic behaviors of rough surfaces.

*Table 5. Comparison of Wenzel and Cassie-Baxter models in surface wettability.*

| | Wenzel Model | Cassie Model |
|---|---|---|
| Water drop profile | The droplet fills all the grooves and valleys of a rough surface. The liquid is in full contact with the surface, enhancing the polymer's intrinsic wettability. | The droplet rests atop the peaks of the roughness with air pockets trapped in the valleys. The rough surface is only partially wet. |
| Apparent contact angle ($\theta^*$) | $\cos(\theta^*) = r.\cos(\theta)$ where r is the roughness factor (ratio of the actual to the projected surface area) and θ is Young's contact angle on a flat surface [48]. | $\cos(\theta^*) = f_1.\cos(\theta_1) + f_2.\cos(\theta_2)$ where $f_1$ and $f_2$ are the respective surface fractions in contact with the liquid and air, and $\theta_1$ and $\theta_2$ are the corresponding contact angles [48]. |
| Notes | - If a surface is intrinsically hydrophobic (θ>90°), roughness will enhance its hydrophobicity.<br>- If it is hydrophilic (θ<90°), roughness will make it even more hydrophilic. | This model is often used to explain the superhydrophobic phenomenon, as seen on lotus leaves. |

For surfaces that combine both nanoscale and micrometer-scale roughness, wetting behavior can be more complex, involving transitions between Wenzel and Cassie states, as sketched in **Figure 16**. Hierarchical structures, which integrate features at several length scales, can enhance certain properties, such as superhydrophobicity [234].

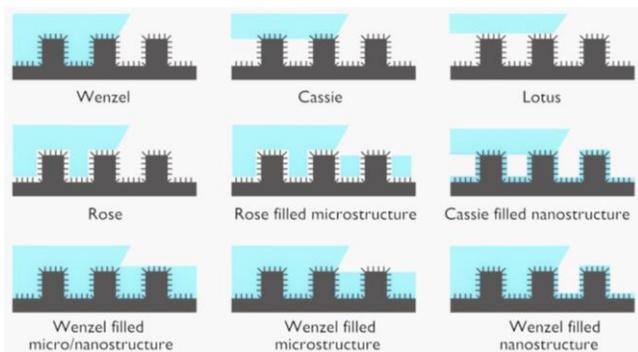

*Figure 16. Schematic of nine wetting scenarios for a surface with hierarchical roughness. Adapted from [234].*

The Lotus wetting state is a classic example of a surface benefiting from hierarchical micro- and nanoscale roughness combined with a hydrophobic coating. The water droplets can hence accumulate and roll off surfaces, such as thermoactive membrane textiles, to eliminate various contaminants [235]. In the rose impregnation state, the water droplet partially wets the surface but does not fully penetrate the microstructures. The droplet resembles a rose petal, with the base in contact with the surface and the top rounded off. This results in strong adhesion, but the contact angle remains relatively high, as evidenced for nanocomposite HMDSO coatings [236]. The scenario in which water partially penetrates the surface microstructures while retaining the "rose" state on a larger scale corresponds to the rose-filled microstructure state. The Cassie-filled nanostructure is a nuanced state in which the liquid droplet is suspended at the top of nanostructures, similar to the Cassie-Baxter model, but specifically applied to nanoscale features. Typically, this state is obtained by covering a micropatterned substrate with electropolymerized nano-structured film [237]. The Wenzel-filled micro/nanostructure is the situation in which the liquid droplet penetrates and completely wets the micro- and nanoscale structures of a surface. The Wenzel-filled microstructure is an extension of the Wenzel state, where the liquid completely penetrates and wets the micro-scale structures on the surface. Conversely, the state where the liquid droplet wets and completely fills the nanostructures on the surface is called a Wenzel-filled nanostructure. These latter impregnation states are not extensively covered in the literature, especially in the context of plasma processing.

The importance of these impregnation states can be illustrated through a concrete application: anti-fog surfaces, the efficiency and durability of which result from a synergy between:
- Surface chemistry: A surface functionalized by oxygen groups can easily spread out water, hence preventing the formation of droplets that blur the view [45].
- Physical structure: A nano-roughened surface traps tiny air pockets, thus influencing the hydrophilic effect [46].

As pointed out by Di Mundo et al., this modification of the surface's physico-chemical properties creates an effective barrier against condensation, inhibiting the formation of water droplets that could obscure vision [238]. **Figure 17** clearly demonstrates this anti-fogging characteristic on polycarbonate (PC) surfaces after plasma treatment. This effect is attributed to the combined impact of surface oxidation (with atomic oxygen concentrations of up to 60%) and nanoscale surface roughness (with a pattern size of 200 to 400 nm). Both factors result in a superhydrophilic PC surface with WCA lower than 10°. Furthermore, while the hydrophilic chemical property may diminish over time, the existing topography helps to maintain the superhydrophilic nature, guaranteeing a long-lasting anti-fog property.

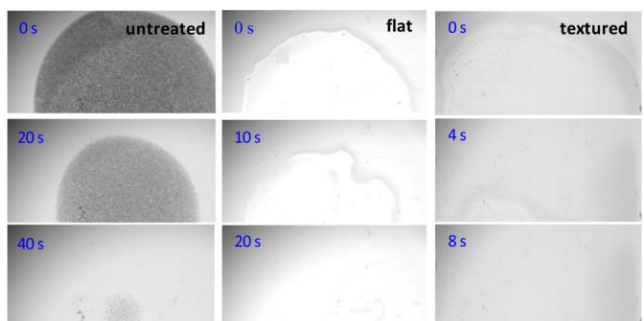

*Figure 17. Time-lapse sequence of condensed water droplets deposited onto transparent polycarbonate (PC) surfaces. (Left column) untreated PC; (middle column) superhydrophilic film obtained after plasma treatment; (right column) superhydrophilic nanotextured film obtained after plasma treatment (Reproduced with permission) [238] Copyright (c) 2014, American Chemical Society.*

### 3.3. Comparing wet chemical approach with dry plasma approach

Both plasma-based and wet-chemical techniques are effective strategies for tailoring the surface properties of polymers, with their respective advantages and disadvantages. While wet-chemical approaches may be more suited to large-scale or cost-sensitive operations, plasma treatments often excel in cases requiring a high degree of precision. The choice between these two methodologies requires an assessment of many elements such as the specific needs of the application, the desired surface characteristics, the type of polymer involved, the scale of production and budgetary considerations. Table 6 provides an in-depth comparison of the comparative advantages and potential limitations of these two methods.

*Table 6. Comparative analysis of plasma processes vs. wet chemical processes in surface treatments.*

| | Plasma Processes | Wet Chemical Processes |
|---|---|---|
| **Uniformity and surface geometry compatibility** | Can uniformly modify surfaces, including those with intricate geometry. | May encounter difficulties ensuring uniform treatment on surfaces with complex geometry due to limitations in chemical access and differential wetting behavior. |
| **Environmental footprint** | Tend to have a reduced environmental footprint as they usually do not require hazardous chemicals and generate minimal waste. | Typically involve solvents and reagents that require stringent disposal procedures to prevent environmental contamination. |
| **Versatility of surface modifications** | Both methods are adaptable, but plasma processes offer a wider range of possible surface activations, cleanings, etchings and graftings in a single step. | |
| **Precision in controlling surface properties** | Allow for fine-tuned control over surface properties by manipulating parameters such as gas type, power, pressure and duration. | May offer less precision in determining final surface properties due to factors such as inconsistent reaction kinetics and diffusion limitations. |
| **Process Speed** | Tend to be relatively rapid, with many procedures only requiring minutes. | Some processes can be more time-consuming, particularly those involving multiple reaction steps or prolonged diffusion times. |
| **Operation temperature** | Can be executed at ambient temperature, making them suitable for heat-sensitive materials. | Some wet chemical treatments might require elevated temperatures. |
| **Post-treatment cleaning requirements** | Generally, do not necessitate post-treatment cleaning as the process leaves no chemical residues. | Often require thorough rinsing or other cleaning procedures to ensure removal of residual reagents and by-products. |

# 4. Plasma-synthesis of polymer films

Chemical vapor deposition (CVD) is a versatile method employed for the fabrication of both organic and inorganic thin films, ranging from monolithic to composite materials. CVD comprises various techniques overviewed in Table 7 and detailed in Section 4.1, namely: thermal CVD (T-CVD), hot filament CVD (HF-CVD), initiated CVD (I-CVD), photo-assisted CVD (PA-CVD), laser-assisted CVD (L-CVD) and plasma-enhanced chemical vapor deposition (PE-CVD). This later technique can be declined in several approaches, detailed in Section 4.2, which correspond to traditional PECVD, aerosol-assisted plasma deposition (AAPD), remote plasma-enhanced chemical vapor deposition (R-PECVD) and pulsed plasma deposition (PPD).

*Table 7. Overview of CVD techniques and approaches, eligible for polymer film deposition.*

| CVD | | | | | | | |
|---|---|---|---|---|---|---|---|
| | | | | | PE-CVD | | |
| HF-CVD | I-CVD | L-CVD | PA-CVD | T-CVD | PPD | AAPD | R-PECVD |

### 4.1. Positioning of PECVD in relation to other CVD techniques

Several chemical vapor deposition (CVD) techniques can be employed for the synthesis of thin polymer films, as summarized in Table 8. T-CVD is a widely-used method where energy is transferred through thermal conduction and radiation from an external heat source (e.g., furnace or hot plate) to both the substrate and precursor gas. This process causes precursor molecules to decompose and form films, as seen in the formation of layered carbon nitride films through melamine powder decomposition [239] or in the gas-phase radical polymerization of di-p-xylylene or di-2-chloro-p-xylylene, resulting in poly-p-xylylene (PPX) and/or poly-2-chloro-p-xylylene (PCPX) films [240]. HF-CVD, a subcategory of T-CVD, utilizes a heated filament near the substrate for localized and controlled energy transfer [241]. This method has been successfully applied to synthesize various thin films, such as fluorocarbon polymer and poly(glycidyl methacrylate) [242, 243]. As an alternative, I-CVD generates free radicals through the thermal decomposition of an initiator, responsible for the polymerization of gaseous monomers. Hence, hydroxypropyl methacrylate vapor and tert-butyl peroxide (TBPO) vapor can be mixed with the latter acting as an initiator [244].

*Table 8. Main characteristics of hot-filament chemical vapor deposition (HF-CVD), initiated chemical vapor deposition (I-CVD), laser chemical vapor deposition (L-CVD), photo-assisted chemical vapor deposition (PA-CVD), plasma-enhanced chemical vapor deposition (PE-CVD) and thermal chemical vapor deposition (T-CVD).*

| | P (mbar) | Gas-Phase Reactions | Precursor Decomposition | Deposition Rate | Deposition Temperature | Film Uniformity | Applications |
|---|---|---|---|---|---|---|---|
| **T-CVD** | 1–1000 | Limited | Surface | Moderate | 300–1200 °C | Good | Semiconductors, dielectrics, metal films |
| **HF-CVD** | 10–1000 | Limited | Surface | Moderate | 300–1000 °C | Good | Thin-film solar cells, amorphous silicon, diamond coatings |
| **I-CVD** | 1–1000 | Yes | Surface and Gas-Phase | Moderate-Fast | 25–150 °C | Good | Polymer films, surface modification, functional coatings |
| **L-CVD** | 1–1000 | Limited | Surface and Gas-Phase | Moderate-Fast | 25–1000 °C | Variable | Patterned films, localized deposition, micro/nano structures |
| **PA-CVD** | 0.1–1000 | Limited | Surface and Gas-Phase | Moderate | 150–700 °C | Good | Hard coatings, diamond-like carbon, wear-resistant films |
| **PE-CVD** | 0.01–1000 | Yes | Surface and Gas-Phase | Moderate-Fast | 25–400 °C | Good | Thin-film transistors, polymers, passivation layers, solar cells |

Moving from thermally driven CVD processes, the techniques that leverage radiation are now considered: L-CVD and PA-CVD, although less commonly used for polymer thin film synthesis. In L-CVD, a focused laser beam imparts energy, either to locally heat the substrate or to dissociate gas-phase precursors such as NH3 and SiCl4, thus enabling the deposition of silicon nitride films [245]. This technique is marked by its high spatial resolution, precision and compatibility with substrates sensitive to high temperatures, which facilitates patterned film deposition and micro/nanostructure fabrication without necessitating supplementary lithography steps. On the other hand, PA-CVD employs either ultraviolet or visible light to trigger chemical reactions, evident in the formation of ultra-thin PMMA films [246]. Apart from providing precise control over the process, PA-CVD operates at lower substrate temperatures, rendering it an ideal choice for materials sensitive to heat.

In the previously discussed CVD techniques, precursor molecules are primarily activated through thermal or radiative processes to grow thin polymer films. Cold plasma can be employed as a disruptive approach, combining these properties with additional factors such as an electrical field, short/long lifespan reactive species and gas flow dynamics to achieve synergistic effects. This approach, known as plasma-enhanced chemical vapor deposition (PECVD) is used to deposit a wide range of materials, including organic materials such as polyethylene [247] and dodecafluoroheptyl methacrylate (ppDFMA) films [248], but also inorganic materials such as silicon dioxide, silicon nitride [249] and titanium dioxide [250].

## 4.2. Main PECVD approaches

The PECVD technique can be carried out over a wide range of plasma sources (see **Figure 4**), operating at different pressure conditions (typically from 0.01 mbar to atmospheric pressure) and for different excitation modes (direct current, alternating current from tens of Hz to GHz, pulses). This multitude of operating parameters gives PECVD the flexibility to meet a variety of process requirements, broadening its potential applications in thin-film deposition.

In the traditional PECVD approach, commonly called "plasma polymerization", the precursors are typically vaporized in a gaseous form. Conversely, the aerosol-assisted plasma deposition (AAPD) technique can be used for precursors that are too sensitive to elevated temperature or that are difficult to vaporize. As a result, various nebulization techniques can be employed, such as a pneumatic nebulizer [251] (pressurized gas, typically air or oxygen) [252, 253], an ultrasonic nebulizer (high-frequency ultrasonic waves) [254], a spray nebulizer [255] (a pump or a pressurized canister to force the liquid through a nozzle) and an electrospray (an electric field disperses a liquid into fine droplets). Regardless of the precursor, the precursor (monomer, polymer or other solution) is first converted into a suspension of tiny liquid droplets within a gas. The resulting aerosol is then introduced into the plasma, where the droplets undergo various chemical reactions before they reach the substrate. As an example, an aqueous dispersion of fluoropolymer (fluoroalkyl acrylate copolymer emulsion in demineralized water) can be directly nebulized into the plasma phase of a DBD, resulting in the deposition of a fluorocarbon nano-layer [252]. Other works include the nebulization of monomers such as hexamethyldisilane [253], heptadeca-fluorodecylacrylate (HDFDA) [251], tetraethylglycoldimethylether (TEGDME) [255] and acrylic acid [254]. Many other examples can be found in [256] for the deposition of thin films that can be organic, inorganic or hybrids.

A second variant of traditional PECVD is remote plasma-enhanced chemical vapor deposition (R-PECVD), where plasma is created at a distance from the substrate, thus reducing potential damage caused by high-energy ions or UV radiation. The method mainly uses gaseous precursors to deposit inorganic films such as silicon dioxide [257] and silicon nitride [258], but it can also be applied to organic films. This technique is favored in the semiconductor sector for its safer, higher-quality deposition of insulating layers than direct PECVD [259].

A third variant of traditional PECVD is pulsed plasma deposition (PPD). This approach provides a fine control of film chemical composition by adjusting the duty cycle of the plasma; i.e., the ratio between the plasma operation time ($\tau_{on}$) and the period ($\tau_{on} + \tau_{off}$). The $\tau_{on}$ phase is marked by the creation of ions (very short-lived) and reactive neutrals (longer-lived), while the $\tau_{off}$ phase sees an increase in the ratio between neutrals and ions. This change in ratio mainly favors film deposition from reactive neutrals, as underlined by d'Agostino and coworkers in the case of vinyltrimethylsilane (VTMS) monomer for the deposition of SiO$_2$-like films [260]. Pulsed plasma chemistry is more sensitive to initial precursor molecules than continuous plasmas, enabling the creation of a wide range of fluorocarbon films from a single precursor.

## 4.3. From molecular precursors in the gas phase to the synthesis of polymer films

The terminology of plasma polymerization (also referred to as glow discharge polymerization or plasma-enhanced polymerization) is somewhat imprecise because the resulting film features a highly branched configuration that deviates from the initial recurring monomer structure [261]. In this section, the terminology of plasma deposition is therefore preferred to plasma polymerization.

Precursor molecules are required to generate the reactive species essential to polymer film formation. They can be introduced into a plasma regardless of their state of fluidity (gas, vapor or liquid), as monomers or not. Monomers, the unitary element required for polymer film synthesis, are small molecules generally carrying reactive functional groups, as already introduced in **Figure 12a**. This reactivity enables the monomers to chemically interact with one another, resulting in the formation of polymer chains. In the plasma phase, these monomers are ionized, fragmented or excited, giving rise to reactive species, which then act as the main instigators of the polymerization process on the substrate surface. Non-monomer precursors, on the other hand, are molecules larger than monomers and include entities such as oligomers and certain inorganic compounds. When exposed to plasma conditions, these larger molecules undergo decomposition or fragmentation to create the reactive species that contribute to the synthesis of the desired film. **Table 9** provides a categorized list of the main precursors used for polymer film growth, distinguishing between monomeric and non-monomeric substances.

*Table 9. Different types of precursors utilized in PECVD: a categorized compilation.*

| Precursors | Categories | Examples |
|---|---|---|
| **Monomers** | Vinyl monomers (Carbon-carbon double bond, C=C) | Methyl methacrylate [113]; styrene [263]; 2-methyl-1,3-butadiene (isoprene) [264]; Vinyltriethoxysilane [265] |
| | Aromatic monomers | Pyrrole [266]; Thiophene [267]; Aniline [268] |
| | Acrylate monomers | 1H,1H,2H,2H-perfluorododecyl acrylate (PFDA) [269]; Dodecylacrylate (DOCA) [269]; Lauryl methacrylate (LMA) [270] |
| | Fluorinated monomers (Carbon-fluorine bonds) | Perfluorinated alkenes, e.g., tetrafluoroethylene (TFE) [271]; Hexafluoropropylene oxide (HFPO) [272]; Hexafluoropropene (HFP) [273]; Perfluorooctyl ethylene [274]; 1H,1H,2H,2H-perfluorooctyl acrylate (PFOA) [275]; Hexafluoroethane (C$_2$F$_6$) [276] |
| | Biocompatible monomers | Allylamine [66,277]; glycidyl methacrylate [243]; acrylic acid [263,278,279]; 2-hydroxyethyl methacrylate (HEMA) [280] |
| **Non-monomers** | Metal-organic compounds | Metal alkoxides, e.g., titanium isopropoxide [251] for titanium dioxide polymerization or tetraethylorthosilicate for silicon dioxide deposition [281] |
| | | Metalorganic complexes, e.g., trimethylaluminum [282] for aluminum oxide thin films or ferrocene for nanostructured hematite thin films [283] |
| | Inorganic compounds | Metal halides, e.g., SiCl$_4$ [284] for silicon deposition |
| | | Metal hydrides, e.g., diborane [285] for boron depositions |
| | Organosilicon compounds | Trimethylsilyl acetate [127] |
| | | Silanes especially tetramethylsilane (TMS) [286], hexamethyldisilazane (HMDSN) [287] |
| | | Siloxanes especially hexamethyldisiloxane (HMDSO) [288] and octamethylcyclotetrasiloxane (OMCTS) [289] |
| | Large organic molecules or oligomers | Perfluorocarbon precursors, e.g., perfluorohexane (PFH, C$_6$F$_{14}$) [290], perfluorodecalin (PFD, C$_{10}$F$_{18}$) [291], perfluoroheptane (PFHp, C$_7$F$_{16}$) [292] |
| | | Ethylene glycol [293], tetra(ethylene glycol) dimethyl ether [294]; Diethylene glycol vinyl ether [295], Diethylene glycol monomethyl ether [296]; ε-caprolactone [296]; Perfluorodecyl acrylate [295] |

The precursor molecules in the plasma phase can generate active species (free radicals, reactive species, etc.) through gaseous reactions such as dissociation, ionization and excitation. In turn, these species can be involved in polymerization reactions as part of the rapid step-growth polymerization (RSGP) mechanism identified by Yasuda. This mechanism, sketched in **Figure 18a**, separates the polymerization reactions into two distinct cycles [270]:

- Cycle I where monovalent reactive species such as free radicals or ions (M$_i$•) interact with the monomers, leading to the formation of covalent bonds and the growth of polymer chains.
- Cycle II where divalent reactive species, i.e., species with two reactive sites (•M$_k$•), interact with the monomers, resulting in the formation of crosslinked or branched polymer structures.

These two cycles are composed of five addition reactions: (1) and (4) occur between a reactive species (M$_i$• or •M$_k$•) and a stable molecule containing a reactive site (M) such as a double or triple bond. Reaction (2) is between two monovalent reactive species, forming an oligomer (M$_i$-M$_j$), which has lost its radical center. Reaction (5) is between two divalent reactive species (•M$_k$• and •M$_j$•) that lead to the formation of a new bond while leaving additional reactive sites for further polymerization. Finally, reaction (3) is referred to as a cross-cycle reaction since it combines single reactive species with divalent reactive ones. This reaction forms a new product (•M$_k$-M$_i$) the radical center of which shows the continuity of the polymerization reaction [296].

The products formed by reactions (1), (3), (4) and (5) can undergo the respective propagation reactions (1'), (3'), (4') and (5'). While (2) would be a termination reaction in conventional polymerization, it can be self-sustained through the collisions with the free electrons from plasma, as represented by reaction (2'). As a whole, the RSGP mechanism corresponds to a succession of termination reactions followed by the reactivation of their products. Repeating these reactions through the two cycles causes the size of the gaseous species to increase and the saturation vapor pressure of the species to decrease. This forces the species to deposit on the substrate as a growing polymer film.

While the RSGP mechanism primarily focuses on the chemical reactions and pathways that lead to the formation of polymer films during the plasma polymerization process, it does not specifically address how the active species interact with the monomer molecules on the substrate, nor does it address the existence of a counter phenomenon likely to reduce or delay film growing efficiency. These two limitations are addressed in another model proposed by Yasuda known as the competitive ablation polymerization (CAP) mechanism [25, 297]. This model, represented in **Figure 18b**, relies on two distinct polymerization paths called plasma-induced polymerization and plasma-state polymerization that can be generalized as plasma-induced deposition and plasma-state deposition:

- The plasma-induced deposition path corresponds to the conventional molecular polymerization of the substrate that is directly exposed to the plasma. The monomers must contain polymerizable structures (e.g., vinyl groups, double bonds, triple bonds, cyclic structures) so that the active species from the plasma can break the double bonds or other reactive sites present in the monomers, hence creating free radicals to initiate polymerization reaction [298]. Then, this reaction propagates on the substrate to gradually form and grow polymer films (see **Figure 18b**). As an example, plasma-induced polymerization from a liquid phase precursor (allyl-substituted cyclic carbonate, A6CC) can drive the synthesis of films bearing functional pendant cyclic carbonates [299]. In addition, the plasma-induced deposition path can be activated at atmospheric pressure; for example, when the plasma treats a layer of liquid monomer (silsesquioxane) covering a porous substrate. The result is the deposition of a highly selective nanocomposite membrane for gas separation [300].

- The plasma-state deposition path involves the formation of polymer-forming intermediate species within the plasma, which then deposit onto the substrate to gradually deposit the polymer film. This path is not based on conventional molecular polymerization but rather on the unique environment and reactive species generated within the plasma [296].

- As illustrated in **Figure 18b**, silane ($SiH_4$) can be used as a starting material, leading to various gas products such as $H_2$ and $SiH_x$ (where x = 1, 2 or 3), while some possible film-forming intermediates can include $Si(OH)_4$ but also $SiH_3$• (silyl radical), $SiH_2$• (disilanyl radical) and $SiH$• (trisilanyl radical). Typically, the deposited film is a silicon-based material that is not necessarily a silicon-based polymer, especially if the resulting plasma-film is a hydrogenated amorphous silicon layer (a-Si:H). Another notable example is HMDSO: a monomer widely used to deposit PDMS-like films on various substrates (e.g., UHMWPE, PTFE, glass) to regulate their wettability properties [153]. Exposure of this monomer in the plasma phase gives rise to various polymer-forming intermediates species, typically •$(CH_3)_2SiO$• species, following the reaction pathway suggested by Hegemann et al. and reported in **Figure 19** [301]. This mechanism involves several reactions such as dissociation, dissociative ionization and dissociative recombination, which are articulated around electron impact threshold energies that typically lie between 3 and 16 eV.

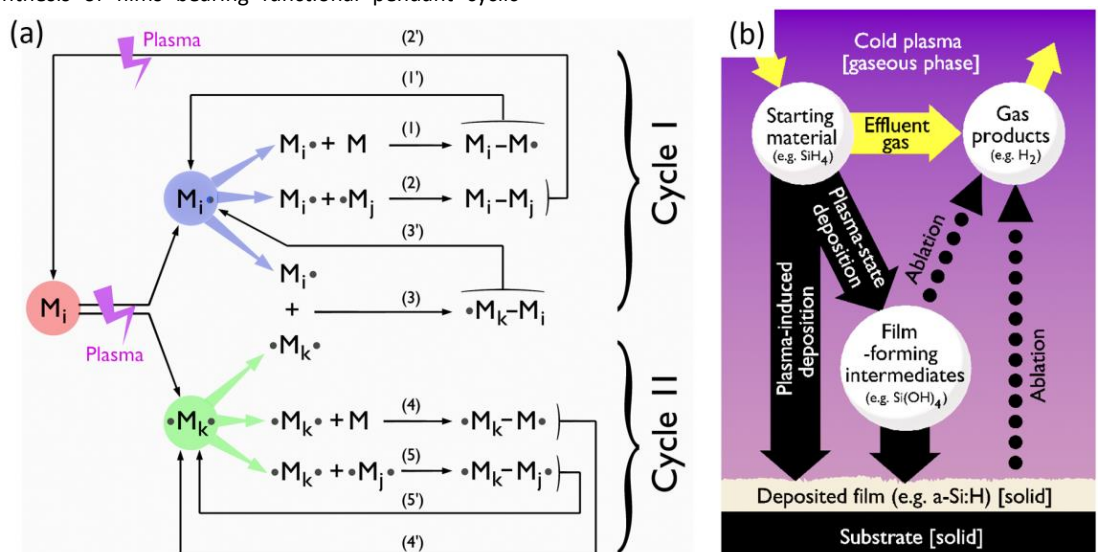

**Figure 18.** *(a) Illustration of Yasuda's Rapid Step-Growth Polymerization (RSGP) mechanism involving two cycles: Cycle I, involving monovalent reactive species and monomers, and Cycle II, with divalent reactive species and monomers, leading to varied polymer structures, adapted from [270]. (b) Depiction of Yasuda's Competitive Ablation Polymerization (CAP) mechanism, detailing the plasma-induced deposition (or plasma-induced polymerization) on substrates and the plasma-state deposition (or plasma-state polymerization), where polymer-forming intermediates are generated within the plasma before depositing onto the substrate. Adapted from [298].*

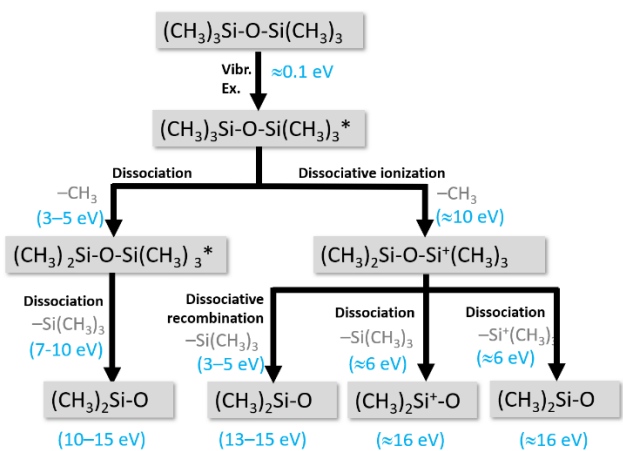

**Figure 19.** *Reaction pathway yielding the generation of film-forming species in HMDSO plasmas (\* represents excited species). Left path is for dissociation while right path is for dissociative ionization with recombination or further dissociation* [301].

If the CAP mechanism explains film growth following two distinct deposition paths, as sketched in **Figure 18b**, it also includes surface etching due to the ion bombardment and eventually VUV radiation. Such etching affects both the substrate and the growing polymer film. The dominating process, whether deposition or etching, depends on various factors; e.g., gas, precursors, substrates and processing conditions [302]. Hence, for pp-fluorinated films, Coburn and Winter deciphered how the substrate DC voltage and F/C ratio modulate the equilibrium between etching and polymerization mechanisms (**Figure 6**, in **Section 2.3.2**). Complementary works from D'Agostino et al. underlined the effects of pressure, Tsubstrate, positive ions and radicals on the deposition mechanisms while both O and F atoms contribute to the etching process [303]. Such mechanisms are deciphered using actinometric optical emission spectroscopy (N₂, Ar, He) [304].

If the RSGP and CAP mechanisms complement each other to decipher the mechanisms upstream of the growth of polymer films, they do not return any information bridging the energy efficiency of the plasma process with the polymer deposition rate. The Yasuda parameter (Y) provides this information as defined in Equation (3):

$$Y = W / (F \cdot M) \qquad (3)$$

where W is the power of the plasma discharge (J·s⁻¹), F is the monomer flow rate (mol·s⁻¹) and M is the molecular weight of the monomer (g·mol⁻¹). As a result, the Yasuda parameter corresponds to the energy input per gram of the polymerized film. Plotting the deposition rate of a polymer film as a function of Y permits the distinguishing of four regimes, represented in **Figure 20a** [305]:

- In the sufficient regime, the energy input per monomer is relatively low. This can lead to the formation of plasma polymer films with lower degrees of crosslinking, less structural stability and potentially more chemical functionality due to a higher concentration of unreacted functional groups. The resulting polymer films may also exhibit more linear chain-like structures and be more susceptible to degradation or environmental factors.

- In the competition regime, the deposition rate remains constant while the Y parameter is changed, whether by increasing P or decreasing F. For styrene, this regime corresponds to a single angular point that bridges monomer-sufficient and -deficient regimes, as sketched in **Figure 20b** [306]. Conversely, this regime exists over a large range of Y values for hydrocarbon monomers such as C₃F₆ and C₄F₁₀ [307].
- The deficient regime indicates a more energy-efficient process, which can result in the formation of polymer films with more desirable properties such as higher crosslink density, better stability and stronger adhesion.
- The atomic deposition process, triggered by high discharge power or low monomer flow rates, results in the extensive fragmentation of monomer molecules, often down to single atoms. This high energy input initiates a series of first-order processes where fragments rearrange, reactivate and further fragment, leading to low deposition rates. Consequently, the composition and structure of the deposited thin film significantly deviate from the original monomer [278].

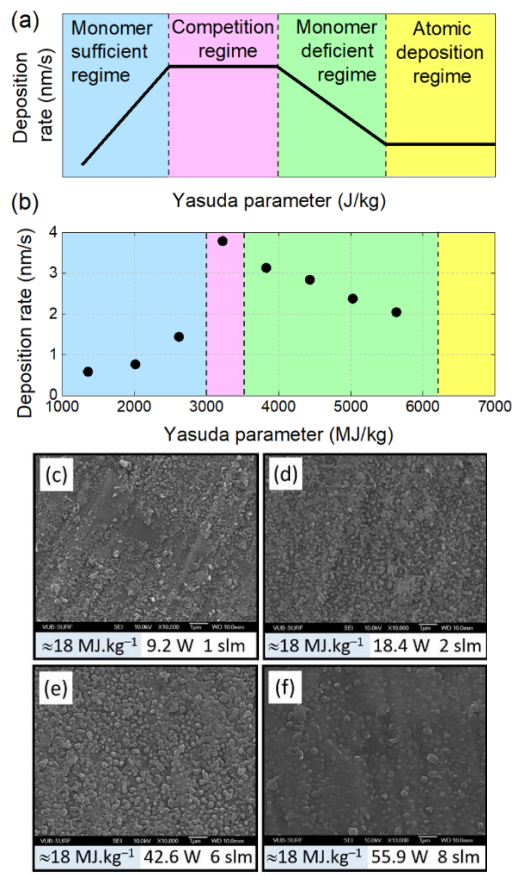

**Figure 20.** *(a) Model of polymer deposition rate vs. Yasuda parameter; (b) Experimental result showing the deposition rate of a plasma-PS film as a function of the Yasuda parameter (styrene density = 0.91 g·cm⁻³, molar mass = 104.15 g·mol⁻¹, $\Phi_{Styrene}$ = 30 µL·min⁻¹ = 4.5 × 10⁻⁴ mol·s⁻¹, P = 7.5 W). Adapted from [278] and [306]. (c–f) Atmospheric plasma coatings of allyl methacrylate observed by SEM (×10 000) for different plasma powers and monomer flow rates while keeping Yasuda parameter fixed at 18 MJ·kg⁻¹. (c) 9.2 W, 1 slm, (d) 18.4 W, 2 slm, (e) 42.6 W, 6 slm, (f) 55.9 W, 8 slm (reproduced with permission)* [308].

In atmospheric pressure sources such as DBD, amplifying the plasma power (achieved by increasing voltage, current or frequency) spatially and temporally influences the properties of the micro-discharges (such as their number, magnitude and individual electrical charge). This, in turn, non-linearly alters the energy distribution among gaseous species (molecular precursors, metastable and reactive species of the plasma). This implies that even when the discharge power and monomer flow rate are proportionally adjusted to keep the Yasuda parameter constant, varied film morphologies and chemistries can emerge. **Figure 20c-f** exemplifies this by demonstrating the morphological alterations of allyl methacrylate (AMA) films deposited by DBD with Y = 18 MJ·kg$^{-1}$ while changing the W and F values [308].

## 4.4. Precursor selection: a pivotal factor in determining semiconductor film properties

The evolution of dielectric materials in the semiconductor industry, particularly for ultra-large-scale integration (ULSI) devices, highlights the importance of precursor selection in defining desired film properties.

Starting with diamond-like carbon (DLC), precursor selection has played a key role in manipulating dielectric constants [309]. The adoption of PECVD has enabled dielectric properties to be controlled. The eventual transition to SiCOH films was also influenced by the choice of precursor, with cyclic organosiloxanes such as octamethylcyclotetrasiloxane (OMCTS) proving optimal [310]. Furthermore, the development of porous SiCOH (pSiCOH) and the use of branched diethoxymethylsilane (DEMS) highlight nuanced changes in precursor selection to achieve specific film attributes.

With regard to the challenges posed by the integration of ultra-low k (ULK) dielectrics, the issue of process-induced damage (PID) is worth highlighting. Processing-induced damage, which affects the film's dielectric constant (k-value), has led to research into precursor modifications. One strategic approach has been to replace certain Si-O-Si bonds with carbosilane (Si-CH$_2$-Si) bonds, which not only retains the desired dielectric properties, but also improves resistance to processing-induced damage [311]. Furthermore, the correlation between PID and the porosity/carbon ratio, combined with the potential ramifications of modifying the precursor blend, highlights the complex relationship between precursor composition and film integrity.

Both perspectives converge on the imperative of striking a delicate balance. There is a constant interplay between mechanical robustness, desired dielectric constants and resistance to process-related damage, all deeply rooted in precursor selection [309]. Thus, precursor selection appears not just as a chemical consideration, but as an essential factor in microprocessor manufacturability, scalability and long-term reliability.

# 5. Technological transfer and future research avenues

## 5.1. Motivations for processing polymers with cold plasma

Although cold plasma processes require specialized expertise and high-cost equipment and produce plasma-polymerized films of limited thickness (from a few nanometers to a few micrometers), they offer substantial advantages. Their inherent flexibility, non-invasive nature, scalability and environmental friendliness make them particularly suitable for polymer surface modification and film deposition, as detailed hereafter.

First, cold plasma technology offers a flexible approach to manipulating the surface properties of polymers. It has the unique ability to impart diametrically opposite properties to the same polymer, depending on the specific experimental conditions applied. To illustrate this specificity, the example of PET is appropriate, as this polymer can be conveniently designed to improve its hydrophilicity or hydrophobicity [312]. On one hand, the atomic oxygen (O) and ozone (O$_3$) species from an oxygen plasma can activate the surface of PET films by introducing hydroxyl (-OH) or carbonyl (C=O) groups [313]. This plasma treatment increases the polymer surface energy and therefore its adhesion properties for applications such as printing, adhesion and coating [314]. On the other hand, the same PET polymer can be exposed to a plasma containing hydrophobic gases such as hexamethyldisiloxane (HMDSO) or perfluorobutane (PFB). The reactive species, namely methyl (-CH$_3$) or fluorine (F), react with the polymer surface to form a thin, conformal coating of hydrophobic groups, for applications such as anti-fouling, water-repellent and oil-resistant coatings [315, 316].

A second compelling benefit of cold plasma processes is their surface-specific nature (also referred to as non-invasive nature), denoting their ability to precisely modify only surface properties without affecting the intrinsic mass characteristics of the polymer [63]. This feature is demonstrated in a study on poly(lactic acid) samples, where DSC measurements indicate that melting enthalpy is unchanged before/after plasma exposure while a decrease in surface wettability is detected by drop shape analysis [317]. Corroborating these observations, Louzi et al. reported similar findings on synthetic polymeric monofilaments such as PP, PET and polyamide-6 when subjected to a corona discharge. The core thermal properties, encompassing melting/crystallization temperatures and the lack of glass transition, remain consistent, while an enhancement in surface wettability is noted [318]. More recently, Fourier-transform infrared (FTIR) spectroscopy has been used to show how air plasma can affix C=O and O–H functional groups onto the surface of PP films, with minimal changes to their bulk properties [319]. While this surface-specific (or non-invasive) nature is observed in most polymers with substantial surface-to-volume ratios (e.g., nanofibers, powders, porous polymers [320, 321, 322]), some exceptions may not be eluded [323].

A third key factor sustaining the suitability of cold plasma for polymer processing is its scalability; i.e., the technology can be easily scaled up to cost-effectively treat larger polymer areas. This is achievable through traditional but efficient roll-to-roll plasma devices characterized by their simplistic design [324]. These devices allow continuous unwinding of the polymer film at a determined speed, ensuring its exposure to the plasma for a determined duration. For example, Stepanova et al. designed a device where polyethylene/polyamide films pass through a region of plasma measuring 8 cm × 20 cm, at speeds between 4 and 16 cm/s, resulting in a plasma exposure time of less than one second [325]. Diffuse coplanar surface barrier discharges (DCDBD) are often implemented in these roll-to-roll devices to generate a high-power density plasma operating in ambient air [326, 327, 328, 329]. In addition to roll-to-roll systems, large-area dielectric devices can be designed for treating large-area polymer, such as automotive exterior parts or large plastic sheets used in construction or signage, whether to solve adhesion or corrosion issues [330]. In contrast to this upscaling approach, APPJs can be employed to treat smaller-scaled polymer parts. Such down-scaled devices are commonly investigated in academic laboratories to specifically address the issues related in Section 3.2. (e.g., surface activation, cleaning, etching, functionalization, etc.) [331, 332, 333, 334].

Finally, cold plasma technology offers a greener and safer alternative to conventional surface modification methods, both by avoiding use and then the release of harmful chemicals into the environment [312]. Cold plasma typically relies on the ionization of a noble gas (e.g., argon) optionally supplemented with a reactive gas (e.g., water vapor). In parallel, some teams directly utilize the ambient air to generate cold plasma [335]. Regardless of the plasma process, the species that are typically generated are reactive oxygen species with a short lifespan such as O and OH radicals (<1 ms) and a long lifespan such as $O_3$ (a few seconds) [336]. This dry approach can effectively activate polymers such as PET, by introducing oxygen-containing functional groups that subsequently increase wettability and adhesion. Once the plasma source is switched off, all the reactive species are recombined after a few seconds. Conversely, traditional wet techniques often utilize a wide array of chemical reagents, including (but not limited to) concentrated sulfuric acid, chromic acid, sodium hydroxide, nickel sulfate and sodium borohydride solutions [337, 338]. Despite their effectiveness, these substances pose considerable environmental risks.

## 5.2. Bridging the gap: current and future prospects for cold plasma applications

The possibilities for future applications of cold plasma-treated polymers in the biomedical field are vast. One could envisage the creation of intelligent drug delivery systems using cold plasma techniques, where polymer capsules could react to specific biological triggers, such as changes in pH, temperature or biomolecular presence, accordingly releasing their drug load [339]. This could enable the precision-targeted treatment of fluctuating conditions such as diabetes or certain types of cancer. Furthermore, we could see the development of biodegradable polymer implants that double as drug delivery systems. Implanted after surgery, these devices could slowly degrade, helping to alleviate pain or prevent infection. Finally, by combining advances in biosensors with cold plasma techniques, we could see the creation of implantable devices capable of real-time health monitoring and drug delivery. For example, a device that continuously measures blood glucose levels and autonomously releases insulin when needed could revolutionize diabetes management. Although these ideas are speculative, they illustrate the innovative potential of cold plasma-treated polymers for the future of medicine.

Membrane-based filtration systems are essential in many fields, including water purification, gas separation and power generation. The performance of these systems largely depends on the selective permeability of the membranes, which determines which substances can pass through [340]. Cold plasma treatments could be used to modify the surface properties of membranes, thereby improving their selectivity and performance. While cold plasmas can be used to modify the pore size and surface chemistry of membranes used for gas separation [341], they could also be useful for developing advanced nanofiltration membranes. These membranes could be used to remove nanoparticles, viruses and other microscopic contaminants from water and air, offering superior purification capabilities. In addition, cold plasma treatment could potentially be used to develop biofiltration systems that selectively remove or deactivate specific pathogens or toxins. This could be particularly useful in healthcare or in the food industry to ensure sterility and safety. Finally, in energy-related applications, plasma-treated membranes could be used in fuel cells, where selective proton or ion transport is essential to their operation. By modifying the membrane's surface properties, it may be possible to optimize ion transport, thereby improving fuel cell efficiency.

Bioplastics, derived from renewable resources such as vegetable fats and oils and corn starch, offer a more sustainable alternative to petroleum-based plastics [342]. However, their application is limited due to inferior mechanical strength, thermal stability and barrier properties [343]. Cold plasma treatment could overcome these difficulties. The plasma process can induce the crosslinking of polymer chains, improving bioplastics' resistance to wear, scratches and other forms of mechanical stress. It could also improve the thermal stability of the material, as the energy required to break these crosslinks can be considerably higher than the energy required to break the original polymer bonds. In addition, cold plasma could offer a means of adjusting the biodegradability of these materials. By tailoring surface properties, it may be possible to influence the material's interactions with the environment and microbes that lead to biodegradation, thereby managing the lifespan of bioplastic products.

Self-healing materials are a class of smart materials that have the inbuilt ability to repair damage caused by mechanical usage over time. This ability to autonomously and inherently repair themselves can extend the life and improve the safety of materials used in a wide variety of applications [344, 345]. In the case of polymers treated or synthesized by cold plasma, this could potentially mean creating a material that can restore its original structure and properties after undergoing mechanical stress, cracks or breaks. Indeed, plasma processes could potentially create the shells of microcapsules containing a healing agent. This could allow for precise control over the properties of the capsule shells, such as their thickness, permeability or rupture strength. Then, plasma treatment could be used to strengthen the adhesion or interaction between the healing agent and the polymer matrix (microcapsules). Plasma processes could also participate in the synthesis of novel healing agents with desired properties.

Intrinsic self-healing materials are polymers that possess the ability to heal or repair themselves without the need for an external healing agent. These materials typically rely on reversible chemical or physical interactions, such as hydrogen bonding or Diels-Alder reactions, to restore their structure and properties after damage [346]. One class of intrinsic self-healing polymers that has attracted interest are those that utilize reversible covalent bonds, such as those formed by Diels-Alder reactions, for self-healing. In this case, polymer chains are crosslinked by reversible Diels-Alder adducts that can break and reform, enabling self-repair. Plasma treatments could be used to refine the surface properties of these polymers. For example, cold plasma treatments could be used to selectively modify the polymer surface to improve its interactions with a specific environment or to introduce additional functionalities. By carefully selecting the gas used in the plasma treatment (e.g., oxygen, nitrogen, etc.), the polymer's surface chemistry could be tailored. In addition, cold plasma treatments could selectively break reversible bonds on the polymer surface, triggering a healing reaction. This could be used, for example, in a controlled way to "refresh" the material surface, thus improving its lifetime.

# 6. Conclusion

Since its inception in the 1950s, plasma technology has undergone a substantial evolution, covering diverse fields from electronics and microelectronics to biomedicine. This evolution, marked by a wave of continuous innovation, has led to notable advances in areas such as thin-film production, fiber processing and, more recently, improved wettability, adhesion and biocompatibility. It is important to note that this technology is geared towards sustainable development, notably by improving the properties of biopolymers.

The complexity of using cold plasma to treat polymers requires a complex understanding of many factors, including the reduced electric field, plasma gas composition, ion energy, UV radiation and temperature. These factors influence the efficiency and characteristics of plasma-treated polymers, making them essential to the successful use of plasma technology in polymer synthesis and modification. Plasma processing is distinguished by its versatility and safety, as it offers uniform methods for modifying polymer surfaces. The effects of these modifications, such as etching, cleaning, decontamination and topographical and chemical changes, expand the applications of polymer materials in unprecedented ways. Furthermore, PECVD facilitates the efficient synthesis of polymer films, thus underlining the need for further research into the delicate balance between deposition and etching, which influences the quality of the films obtained.

Cold plasma technology has proven to be a game-changer in polymer processing. Even with its current limitations, its potential to shape the future of biomedicine, filtration systems, bioplastics and self-healing materials is promising. More focused research is therefore needed to fully exploit the ability of plasma technology to control surface properties. The future of plasma technology in polymer processing is therefore promising, with a trajectory of continued growth and refinement. With its proven flexibility, ability to modify specific surfaces, scalability and environmental sustainability, cold plasma technology could be a transformative force in materials science and its real-world applications. Further research into this technology could lead to revolutionary advances in intelligent drug delivery systems, advanced purification membranes, stronger bioplastics and self-healing materials. This study underlines the importance of ongoing research into plasma technology for the betterment of our society.